\documentclass[12pt]{article}
\usepackage[margin=1in]{geometry}
\usepackage{setspace}
\usepackage{titling}
\usepackage{color}
\usepackage{graphicx}
\usepackage[bf,margin=20pt,format=hang,font={small,stretch=1.1}]{caption}
\usepackage{hyperref}
\usepackage{amsmath,amssymb}
\usepackage{tikz}

\newcommand\TL{\hfil$\displaystyle{##}$}
\newcommand\TR{$\displaystyle{{}##}$\hfil}

\newcommand\TT{\hbox{##}}
\def\seqalign#1#2{\vcenter{\openup1\jot
  \halign{\strut #1\cr #2 \cr}}}
\def\lbldef#1#2{\expandafter\gdef\csname #1\endcsname {#2}}
\newcommand{\eqn}[3][]{\lbldef{#2}{(\ref{#2})}%
\begin{equation} \eqalign{#3} \label{#2} \end{equation}}
\def\eqalign#1{\vcenter{\openup1\jot
    \halign{\strut\span\TL & \span\TR\cr #1 \cr
   }}}
\def\eno#1{(\ref{#1})}
\def\href#1#2{#2}

\newcommand{\SimpleFigure}[2][3in]{%
\IfFileExists{figures/#2.pdf}{\includegraphics[width=#1]{figures/#2.pdf}}{%
\IfFileExists{figures/#2.pdf}{\includegraphics[width=#1]{figures/#2.jpg}}{%
\IfFileExists{figures/#2.png}{\includegraphics[width=#1]{figures/#2.png}}{%
\includegraphics[width=#1]{figures/UnderConstruction.jpg}}}}}
\newcommand{\CaptionedFigure}[3][3in]{%
\begin{figure}\centerline{%
\IfFileExists{figures/#2.pdf}{\includegraphics[width=#1]{figures/#2.pdf}}{%
\IfFileExists{figures/#2.jpg}{\includegraphics[width=#1]{figures/#2.jpg}}{%
\IfFileExists{figures/#2.png}{\includegraphics[width=#1]{figures/#2.png}}{%
\includegraphics[width=#1]{figures/UnderConstruction.jpg}}}}}%
\caption{\if\relax\detokenize{#3}\relax Caption for {\tt #2}\else #3\fi}\label{#2}%
\end{figure}}
\makeatletter\let\over\@@over\makeatother

\renewenvironment{abstract}
 {\normalsize
  \begin{center}
   \bfseries \abstractname\vspace{-.5em}\vspace{0pt}
  \end{center}
  \list{}{
   \setlength{\leftmargin}{0in}%
   \setlength{\rightmargin}{\leftmargin}%
  }%
  \item\relax}
 {\endlist}

\title{Continuum limits of sparse coupling patterns}

\author{Steven S.~Gubser, Christian Jepsen, Ziming Ji, and Brian Trundy}
\date{}

\begin{document}
\begin{titlingpage}

\setlength{\droptitle}{-70pt}
\maketitle
\begin{abstract}
We exhibit simple lattice systems, motivated by recently proposed cold atom experiments, whose continuum limits interpolate between real and $p$-adic smoothness as a spectral exponent is varied.  A real spatial dimension emerges in the continuum limit if the spectral exponent is negative, while a $p$-adic extra dimension emerges if the spectral exponent is positive.  We demonstrate H\"older continuity conditions, both in momentum space and in position space, which quantify how smooth or ragged the two-point Green's function is as a function of the spectral exponent.  The underlying discrete dynamics of our model is defined in terms of a Gaussian partition function as a classical statistical mechanical lattice model.  The couplings between lattice sites are sparse in the sense that as the number of sites becomes large, a vanishing fraction of them couple to one another.  This sparseness property is useful for possible experimental realizations of related systems.
\end{abstract}
\vfill
May 2018
\end{titlingpage}

\tableofcontents

\clearpage

\section{Introduction}

The $p$-adic numbers (for any fixed choice of a prime number $p$) are an alternative way of filling in the ``gaps'' between rational numbers in order to form a complete set, or continuum.  They have been studied for over a hundred years, and one of many mathematical introductions to the subject is \cite{Gouvea:1997zz}.  The key ingredient is the $p$-adic norm.  Briefly, one defines $|a|_p \equiv p^{-v(a)}$ for non-zero $a \in \mathbb{Z}$, where $v(a)$ (the so-called valuation of $a$) is the number of times $p$ divides $a$.  Then $|a/b|_p \equiv p^{-v(a)+v(b)}$ for non-zero integers $a$ and $b$.  By fiat, $|0|_p = 0$.  This norm is very different from the usual absolute value, which is denoted $|x|_\infty$ to avoid any ambiguity.  Just as the real numbers $\mathbb{R}$ are the completion of the rationals $\mathbb{Q}$ with respect to $|\cdot|_\infty$, so the $p$-adic numbers $\mathbb{Q}_p$ are the completion of $\mathbb{Q}$ with respect to $|\cdot|_p$ for any fixed $p$.\footnote{Let's briefly review what completion means.  A Cauchy sequence $\{x_n\}_{n=1}^\infty$ with respect to a norm $|\cdot|$ is one where for any real number $\epsilon>0$, we have $|x_n-x_m| < \epsilon$ provided $n$ and $m$ are larger than some minimum value $N$ (which usually depends on $\epsilon$).  The reals $\mathbb{R}$ can be understood as the set of Cauchy sequences of rational numbers, modulo an equivalence relation defined by considering two Cauchy sequences equivalent iff combining them by alternating terms gives again a Cauchy sequence.  The $p$-adic numbers $\mathbb{Q}_p$ are the completion of $\mathbb{Q}$ with respect to the $p$-adic norm $|\cdot|_p$.  The four field operations, namely addition, subtraction, multiplication, and division by non-zero elements, are defined on $\mathbb{Q}_p$ by continuity from their standard definition on $\mathbb{Q}$.  Complications arise if one attempts to proceed similarly with non-prime $p$: In particular, the obvious Cauchy construction results in a ring, not a field---in fact, a ring in which one can have $xy=0$ with both $x$ and $y$ non-zero.}  We will often describe the real numbers as Archimedean because the norm $|\cdot|_\infty$ satisfies the Archimedean property, namely that if $0 < |a|_\infty < |b|_\infty$, then for some $n \in \mathbb{Z}$ we have $|na|_\infty > |b|_\infty$.  This property fails for the $p$-adic norm because it enjoys instead the so-called ultra-metric inequality, $|a+b|_p \leq \max\{|a|_p,|b|_p\}$, which implies in particular that $|na|_p \leq |a|_p$ for all $n \in \mathbb{Z}$.

Field theories (in the physics sense of ``field'') over the $p$-adic numbers have been studied extensively, starting with Dyson's hierarchical model \cite{Dyson:1968up} and continuing with the rigorous results of \cite{Bleher:1973zz}, with the field theory perspective emerging clearly in \cite{Lerner:1989ty}.  Recent reviews include \cite{Missarov:2012zz}.  The essential features that we will use in this work can already be understood, for $p=2$, in terms of a slight rephrasing of Dyson's original work, as follows.  Consider the ``furthest neighbor'' Ising model.  By this we mean that starting with $2^N$ Ising spins, numbered $0$ through $2^N-1$, we strongly couple the spins which are as far apart as possible as measured through sequential counting.  Thus spin $0$ couples to spin $2^{N-1}$, spin $1$ to spin $2^{N-1}+1$, and so forth.  Arranging the spins on a circle, we are coupling pairs of spins which are diametrically opposite.  If we stopped there, we would have $2^{N-1}$ strongly coupled pairs of spins, with each pair entirely decoupled from every other pair.  We want a more interesting thermodynamic limit, so we keep going by coupling each pair of spins with the pair furthest from it (again in the sense of sequential counting).  Then we couple pairs of pairs, and so on.  At each stage we reduce the coupling strength by a fixed factor $2^{s+1}$, where $s \in \mathbb{R}$ is what we will call the spectral parameter.  This coupling pattern can be expressed concisely in terms of the $2$-adic norm of the separation of the spins, as we will specify in more detail in section~\ref{SPARSE}.  The overall picture is illustrated in figure~\ref{FurthestNeighbor}.
\usetikzlibrary{math}
 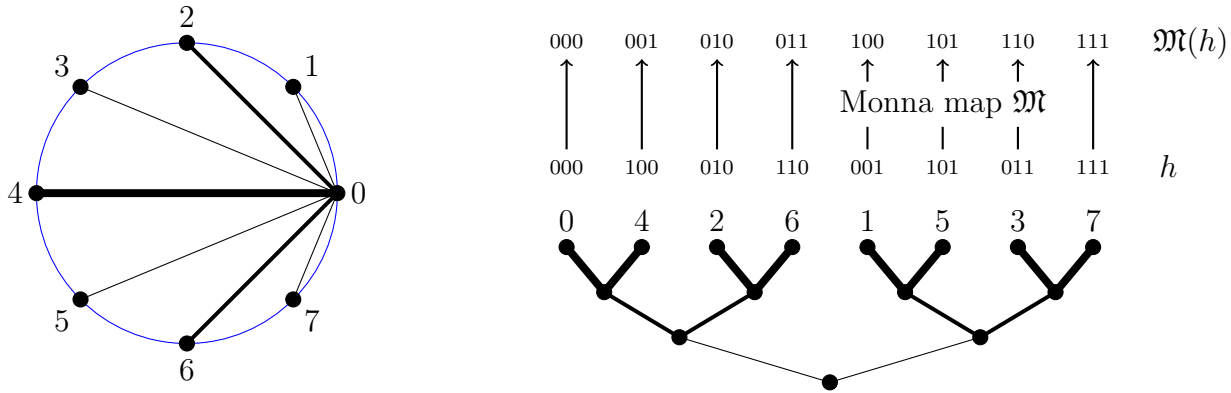
\begin{figure}
  \centerline{\begin{tikzpicture}
   %Define coordinates for the eight points around a circle
   \tikzmath{
     \rad = 2;
     \pointsize = 0.1;
     \x0 = \rad * cos(0); \y0 = \rad * sin(0);
     \x1 = \rad * cos(45); \y1 = \rad * sin(45);
     \x2 = \rad * cos(90); \y2 = \rad * sin(90);
     \x3 = \rad * cos(135); \y3 = \rad * sin(135);
     \x4 = \rad * cos(180); \y4 = \rad * sin(180);
     \x5 = \rad * cos(225); \y5 = \rad * sin(225);
     \x6 = \rad * cos(270); \y6 = \rad * sin(270);
     \x7 = \rad * cos(315); \y7 = \rad * sin(315);
    }
   \coordinate (C) at (0,0);
   \coordinate (R0) at (\x0,\y0);
   \coordinate (R1) at (\x1,\y1);
   \coordinate (R2) at (\x2,\y2);
   \coordinate (R3) at (\x3,\y3);
   \coordinate (R4) at (\x4,\y4);
   \coordinate (R5) at (\x5,\y5);
   \coordinate (R6) at (\x6,\y6);
   \coordinate (R7) at (\x7,\y7);
   %Draw the circle and the points
   \draw [blue] (0,0) circle [radius=\rad];
   \draw [fill] (R0) circle [radius=\pointsize];
   \draw [fill] (R1) circle [radius=\pointsize];
   \draw [fill] (R2) circle [radius=\pointsize];
   \draw [fill] (R3) circle [radius=\pointsize];
   \draw [fill] (R4) circle [radius=\pointsize];
   \draw [fill] (R5) circle [radius=\pointsize];
   \draw [fill] (R6) circle [radius=\pointsize];
   \draw [fill] (R7) circle [radius=\pointsize];
   %Label the points.  right=0.1pt means to the right by the standard length plus 1pt
   \node [right=1pt] at (R0) {0};
   \node [above right] at (R1) {1};
   \node [above=2pt] at (R2) {2};
   \node [above left] at (R3) {3};
   \node [left=1pt] at (R4) {4};
   \node [below left] at (R5) {5};
   \node [below=2pt] at (R6) {6};
   \node [below right] at (R7) {7};
   %Connect the points according to strength of coupling
   \draw [line width=3pt] (R0) -- (R4);
   \draw [line width=1.5pt] (R0) -- (R2);
   \draw [line width=1.5pt] (R0) -- (R6);
   \draw (R0) -- (R1);
   \draw (R0) -- (R3);
   \draw (R0) -- (R5);
   \draw (R0) -- (R7);
  \end{tikzpicture}\hfill
  \begin{tikzpicture}
   \tikzmath{
     \hSpacing = 1;
     \vSpacing = 0.6;
     \pointsize = 0.1;
     \x0 = 0; \y0 = 0;
     \x1 = \hSpacing; \y1 = 0;
     \x2 = 2 * \hSpacing; \y2 = 0;
     \x3 = 3 * \hSpacing; \y3 = 0;
     \x4 = 4 * \hSpacing; \y4 = 0;
     \x5 = 5 * \hSpacing; \y5 = 0;
     \x6 = 6 * \hSpacing; \y6 = 0;
     \x7 = 7 * \hSpacing; \y7 = 0;
    }
   %Top layer
   \coordinate (H0) at (\x0,\y0);
   \coordinate (H1) at (\x1,\y1);
   \coordinate (H2) at (\x2,\y2);
   \coordinate (H3) at (\x3,\y3);
   \coordinate (H4) at (\x4,\y4);
   \coordinate (H5) at (\x5,\y5);
   \coordinate (H6) at (\x6,\y6);
   \coordinate (H7) at (\x7,\y7);
   \draw [fill] (H0) circle [radius=\pointsize];
   \draw [fill] (H1) circle [radius=\pointsize];
   \draw [fill] (H2) circle [radius=\pointsize];
   \draw [fill] (H3) circle [radius=\pointsize];
   \draw [fill] (H4) circle [radius=\pointsize];
   \draw [fill] (H5) circle [radius=\pointsize];
   \draw [fill] (H6) circle [radius=\pointsize];
   \draw [fill] (H7) circle [radius=\pointsize];
   \node [above=2pt] at (H0) {0};
   \node [above=2pt] at (H1) {4};
   \node [above=2pt] at (H2) {2};
   \node [above=2pt] at (H3) {6};
   \node [above=2pt] at (H4) {1};
   \node [above=2pt] at (H5) {5};
   \node [above=2pt] at (H6) {3};
   \node [above=2pt] at (H7) {7};
   %Second layer
   \tikzmath{
    \xS0 = (\x0+\x1)/2; \yS0 = \y0-\vSpacing;
    \xS2 = (\x2+\x3)/2; \yS2 = \y2-\vSpacing;
    \xS4 = (\x4+\x5)/2; \yS4 = \y4-\vSpacing;
    \xS6 = (\x6+\x7)/2; \yS6 = \y6-\vSpacing;
   }
   \coordinate (H01) at (\xS0,\yS0);
   \coordinate (H23) at (\xS2,\yS2);
   \coordinate (H45) at (\xS4,\yS4);
   \coordinate (H67) at (\xS6,\yS6);
   \draw [fill] (H01) circle [radius=\pointsize];
   \draw [fill] (H23) circle [radius=\pointsize];
   \draw [fill] (H45) circle [radius=\pointsize];
   \draw [fill] (H67) circle [radius=\pointsize];
   %Third layer
   \tikzmath{
    \xT0 = (\xS0+\xS2)/2; \yT0 = \yS0-\vSpacing;
    \xT4 = (\xS4+\xS6)/2; \yT4 = \yS4-\vSpacing;
   }
   \coordinate (H0123) at (\xT0,\yT0);
   \coordinate (H4567) at (\xT4,\yT4);
   \draw [fill] (H0123) circle [radius=\pointsize];
   \draw [fill] (H4567) circle [radius=\pointsize];
   %Fourth layer
   \tikzmath{
    \xF0 = (\xT0+\xT4)/2; \yF0 = \yT0-\vSpacing;
   }
   \coordinate (H01234567) at (\xF0,\yF0);
   \draw [fill] (H01234567) circle [radius=\pointsize];
   %Connections between layers
   \draw [line width=3pt] (H0) -- (H01) -- (H1);
   \draw [line width=3pt] (H2) -- (H23) -- (H3);
   \draw [line width=3pt] (H4) -- (H45) -- (H5);
   \draw [line width=3pt] (H6) -- (H67) -- (H7);
   \draw [line width=1.5pt] (H01) -- (H0123) -- (H23);
   \draw [line width=1.5pt] (H45) -- (H4567) -- (H67);
   \draw (H0123) -- (H01234567) -- (H4567);
   %Binary source of Monna
   \tikzmath{
    \yD = 1.3; \yI = 2.5;
    \yM = (\yD+\yI)/2;
    \xh = 8 * \hSpacing;
   }
   \node [below] at (\x0,\yD) {\scriptsize 000};
   \node [below] at (\x1,\yD) {\scriptsize 100};
   \node [below] at (\x2,\yD) {\scriptsize 010};
   \node [below] at (\x3,\yD) {\scriptsize 110};
   \node [below] at (\x4,\yD) {\scriptsize 001};
   \node [below] at (\x5,\yD) {\scriptsize 101};
   \node [below] at (\x6,\yD) {\scriptsize 011};
   \node [below] at (\x7,\yD) {\scriptsize 111};
   \node [above] at (\x0,\yI) {\scriptsize 000};
   \node [above] at (\x1,\yI) {\scriptsize 001};
   \node [above] at (\x2,\yI) {\scriptsize 010};
   \node [above] at (\x3,\yI) {\scriptsize 011};
   \node [above] at (\x4,\yI) {\scriptsize 100};
   \node [above] at (\x5,\yI) {\scriptsize 101};
   \node [above] at (\x6,\yI) {\scriptsize 110};
   \node [above] at (\x7,\yI) {\scriptsize 111};
   \draw [->, thick] (\x0,\yD) -- (\x0,\yI);
   \draw [->, thick] (\x1,\yD) -- (\x1,\yI);
   \draw [->, thick] (\x2,\yD) -- (\x2,\yI);
   \draw [->, thick] (\x3,\yD) -- (\x3,\yI);
   \node at (\x5,\yM) {Monna map $\mathfrak{M}$};
   \draw [thick] (\x4,\yD) -- (\x4,\yM-0.3);
   \draw [->, thick] (\x4,\yM+0.3) -- (\x4,\yI);
   \draw [thick] (\x5,\yD) -- (\x5,\yM-0.3);
   \draw [->, thick] (\x5,\yM+0.3) -- (\x5,\yI);
   \draw [thick] (\x6,\yD) -- (\x6,\yM-0.3);
   \draw [->, thick] (\x6,\yM+0.3) -- (\x6,\yI);
   \draw [->, thick] (\x7,\yD) -- (\x7,\yI);
   \node [below=-2pt] at (\xh,\yD) {$h$};
   \node [above=-4pt] at (\xh+0.3,\yI) {$\mathfrak{M}(h)$};
  \end{tikzpicture}}
  \caption{Left: A furthest neighbor coupling pattern among eight spins.  The thickness of lines indicates the strength of the coupling between spin $0$ and the other spins.  The coupling pattern is invariant under shifting by a lattice spacing, so for example spins $1$ and $5$ are as strongly coupled as spins $0$ and $4$.  The blue circle is to guide the eye and does not indicate additional couplings.\\[3pt]
  Right: A hierarchical representation of the couplings between spins.  Above each spin's label we have given the base $2$ presentation of the spin number, and we have shown how the Monna map acts on these numbers by reversing digits in the base $2$ presentation.}\label{FurthestNeighbor}
 \end{figure}

A natural way to understand the furthest neighbor model is in terms of a hierarchy of clusters of spins, as also illustrated in figure~\ref{FurthestNeighbor}.  The hierarchical tree of these clusters gives a particularly clear understanding of the $2$-adic distance, because if we define $d(i,j)$ as the number of steps required to go from point $i$ to point $j$ on the tree, then for boundary points $i$ and $j$ taking integer values between $0$ and $2^N-1$, we have $|i-j|_2 = 2^{-N+d(i,j)/2}$.

A further step is to send the position $h$ of each spin through the discrete Monna map $\mathfrak{M}$, which takes as its argument an integer between $0$ and $2^N-1$ and returns an integer in the same range obtained by reversing the base $2$ digits of the argument.  This map is intuitively useful because after applying it, positions which were close hierarchically are close sequentially.  Note however that this closeness relationship doesn't work in reverse: positions that are sequentially close need after applying $\mathfrak{M}$ need not have been hierarchically close before applying $\mathfrak{M}$.

Not surprisingly, Green's functions of spins in the furthest neighbor Ising model depend on the locations $i$ and $j$ of the spins only through the $2$-adic norm $|i-j|_2$.  Formally, this is a consequence of invariance of the partition function under relabeling all lattice sites according to $i \to u i + b$ where $u$ and $b$ are elements of $\mathbb{Z}/2^N\mathbb{Z}$ and $|u|_2 = 1$.  Note that the statement $|u|_2=1$ is well defined in $\mathbb{Z}/2^N\mathbb{Z}$: it amounts to requiring $u$ to be an odd number.  Intuitively, $i \to u i + b$ is like a rotation followed by a translation.  Translational invariance means that the Green's function can depend on $i$ and $j$ only through their difference $i-j$.  ``Rotational'' invariance implies that the dependence on $i-j$ must be only through the norm $|i-j|_2$.

The sparse coupling pattern that we want to study eliminates couplings between spins $i$ and $j$ unless $i-j$ is a power of $2$, or minus a power of $2$, modulo $2^N$.  So for the case $N=3$ shown in figure~\ref{FurthestNeighbor}, we drop the coupling between spins $0$ and $3$, and between $0$ and $5$, and between all translated copies of these pairs, for example the pairs $(1,4)$ and $(1,6)$.  For this small value of $N$, obviously the ``sparse'' coupling pattern is still nearly all-to-all.  But for large $N$, the number of spins coupling to spin $0$ increases linearly with $N$ while the total number of spins is $2^N$.  This type of coupling pattern was first brought to our attention in discussions about proposed cold atom experiments \cite{SchleierSmithDiscussions}. 

If the spectral exponent $s$ is large and positive, we expect to recover nearly the same results as if we had used a truly $2$-adic all-to-all coupling as in the furthest neighbor model described previously.  Here's why.  When $s$ is large and positive, the coupling between spins $0$ and $2^{N-1}$ produce very tightly coupled pairs, and the pairs of pairs are also pretty tightly coupled.  This tight coupling means that when we proceed to the next level down the tree and couple $0$ relatively weakly to spins $\pm 2^{N-3}$ but not $\pm 3 \times 2^{N-3}$, all that matters to a good approximation is that we are coupling the quartet $\{0,2^{N-1},\pm 2^{N-2}\}$ to the quartet $\{\pm 2^{N-3},\pm 3 \times 2^{N-3}\}$.  Likewise, as we go further down the hierarchy of couplings, while it's true that we couple spins in previously established $2$-adic clusters unequally, the clusters at each step are so tightly bound within themselves relative to their coupling with each other that they act almost like single spins.

Meanwhile, as we will see, when the spectral parameter $s$ is made large and negative, we recover nearest neighbor interactions.  The two-point Green's function of the nearest neighbor model with $2^N$ spins is then well-approximated at large $N$ by a continuum Green's function that we can extract from field theory over $\mathbb{R}$.  This Green's function is smooth in an Archimedean sense, except at zero separation: In fact, if we are considering the model with pure nearest neighbor interactions, the Green's function away from zero separation is $C^\infty$.  The smoothness of the continuum limit of the Green's function is a good way to understand how continuous quantities emerge from a discrete lattice description.  Poetically, a continuous spatial dimension emerges from nearest neighbor interactions on a large discrete lattice.

A natural question to follow up the discussion of the previous paragraph is, what counts as a smooth continuum Green's function from a $2$-adic point of view?  Let's revert to discussing $p$-adic smoothness for any prime $p$, since it is no more difficult than for $p=2$.  Continuity is easy to understand over the $p$-adic numbers: If a function $G$ maps $\mathbb{Q}_p$ to $\mathbb{R}$, then we can define $G$ as continuous at $x$ if for any $\epsilon>0$ there exists a $\delta>0$ such that every $y$ with $|x-y|_p < \delta$ has $|G(x)-G(y)|_\infty < \epsilon$.  It is harder to find the proper analog of a $C^\infty$ condition on $G$, because derivatives of $G$ with respect to $x$ are tricky to define.  (Heuristically, that's because $dG/dx$ is neither real nor $p$-adic, but apparently some ratio of the two, which doesn't quite make sense.)  In fact, the accepted analog of a $C^\infty$ condition is to require that a map $G$ from $\mathbb{Q}_p$ to $\mathbb{R}$ is locally constant.\footnote{A more complete introduction to smooth test functions over the $p$-adic numbers than we will provide can be found, for example, in \cite{Sally:1998zz}.}  For a function to be locally constant at a point $x$, we must be able to find some $\delta>0$ such that every $y$ with $|x-y|_p < \delta$ has $G(x)=G(y)$.  Surprisingly, a function from $\mathbb{Q}_p$ to $\mathbb{R}$ which is everywhere locally constant need not be globally constant (as it would for a function from $\mathbb{R}$ to $\mathbb{R}$).  For example, the function which is $1$ on $\mathbb{Z}_p$ and $0$ over the rest of $\mathbb{Q}_p$ is locally constant everywhere, but obviously not globally constant.  Green's functions in models with perfectly $p$-adic coupling are also locally constant except at zero separation, as we will see in examples soon.  When we turn to sparse coupling patterns, we will recognize that we are recovering $2$-adic continuity precisely when the two-point Green's function is well approximated by a locally constant function.  This is exactly what happens in the limit of large positive $s$ for the $2$-adic statistical mechanical models that we will study explicitly.

In short, as the spectral exponent $s$ ranges from large negative to large positive values, the Green's functions we study transition from showing emergent Archimedean continuity to showing emergent $p$-adic continuity.  How this transition occurs is slightly subtle, but we will combine some numerical results with analytical reasoning to characterize it both in momentum space and position space.

The sparse coupling pattern we consider was suggested to us in connection with prospective cold atom experiments in which coupling patterns of at least approximately the form we consider may be realized \cite{SchleierSmithDiscussions}, using techniques along the lines of \cite{Hung:2016zz}.  It is outside our present scope to provide a detailed account of these experiments, but let us mention three salient points:
 \begin{enumerate}
  \item Translational invariance of the coupling (except for endpoint effects) is a natural feature of the experimental setup.
  \item While it is possible in principle to arrange a wide variety of couplings, it is useful to focus on sparse couplings, because every time a coupling is introduced between spins at fixed separation, it increases dissipative tendencies in the system.
  \item The most straightforward models to realize in the cold atom system are XXZ Heisenberg models with no on-site terms, i.e.~with hopping terms only.  We will be dealing with a substantially simpler statistical mechanical model in this paper but hope to return to the more complicated dynamics of the XXZ model in future work.
 \end{enumerate}

The organization of the rest of this paper is as follows.  In section~\ref{SETUP} we describe the class of statistical mechanical, one-dimensional spins chains that we will study, and we give a general account of how to compute Green's functions before treating in turn four models within this class: Nearest neighbor interactions, power-law interactions, $p$-adic interactions (in principle for any prime $p$ though we eventually focus on $p=2$), and finally sparse couplings, which interpolate between nearest neighbor and $p$-adic behavior.  In section~\ref{CONTINUITY}, starting from field theory, we obtain H\"older continuity bounds on the continuum limit of Green's functions computed in section~\ref{SETUP}.  In section~\ref{NUMERICS}, we show through numerical studies that the smoothness of Green's functions in momentum space is well captured by the H\"older continuity bounds derived in section~\ref{CONTINUITY}.  Position space continuity is more complex, with different H\"older exponents depending on whether one is looking at global or local smoothness properties.

\section{The statistical mechanical models of interest}
\label{SETUP}

Our aim is to work out the statistical mechanics of models with a variety of non-local couplings.  We want our results to be as explicit as possible, and to have as few parameters as we can arrange.  Consider therefore the following Hamiltonian for a lattice with $L$ sites:
 \eqn{genH}{
  H \equiv -{1 \over 2} \sum_{i,j} J_{ij} \phi_i \phi_j - \sum_j h_j \phi_j \,,
 }
where the $\phi_i$ are still commuting real numbers.  Clearly $J_{ij} = J_{ji}$ because $\phi_i \phi_j$ is symmetric.  Let us also assume translational invariance: That is, $J_{ij} = J_{i-j}$, where arithmetic operations like $i-j$ are carried out modulo $L$.  Define $L$-dimensional vectors $\vec{v}_\kappa$ by
 \eqn{vFourier}{
  v_{\kappa,j} \equiv {1 \over \sqrt{L}} e^{2\pi i \kappa j / L} \qquad\hbox{for}\qquad
   \kappa = 0,1,2,\dots,L-1 \,.
 }
Any quantity $X_j$ depending on $j \in \{0,1,2,\dots,L-1\}$ can be Fourier transformed according to
 \eqn{Xfourier}{
  X_j = \sum_{\kappa=0}^{L-1} \tilde{X}_\kappa v_{\kappa,j} \,.
 }
An easy calculation shows that
 \eqn{Jefs}{
  {\bf J} \vec{v}_\kappa = \sqrt{L} \tilde{J}_\kappa \vec{v}_\kappa \,,
 }
where ${\bf J}$ without indices means the symmetric matrix $J_{ij}$, and $\tilde{J}_\kappa$ is the Fourier transform of the coupling strengths $J_h$.  Using \eno{Xfourier}-\eno{Jefs}, we have immediately
 \eqn{Hdiag}{
  H = -{\sqrt{L} \over 2} \sum_{\kappa=0}^{L-1} \tilde{J}_\kappa \tilde\phi_{-\kappa} 
   \tilde\phi_\kappa - \sum_{\kappa=0}^{L-1} \tilde{h}_{-\kappa} \phi_\kappa \,.
 }
We now make two assumptions:
 \begin{itemize}
  \item $\tilde{J}_0=0$.  We understand this as a consequence of assuming the existence of a symmetry where all the $\phi_i$ are shifted by a common value.
  \item $\tilde{J}_\kappa < 0$ for all $\kappa \neq 0$.  This amounts to saying that the interactions among the $\phi_i$ are ferromagnetic.
 \end{itemize}
It is useful to note that the second assumption follows from the first together with the requirement that all $J_h \geq 0$ for $h \neq 0$, with not all of them equal to zero.

In order to make the statistical mechanics of $H$ well-defined, we insert a factor of $\delta(\tilde\phi_0)$ into the partition function:
 \eqn{Zgen}{
  Z[h] \equiv \left( \prod_{j=0}^{L-1} \int_{-\infty}^\infty d\phi_j \right) \delta(\tilde\phi_0)
    e^{-\beta H} = Z[0] \exp\left\{ -{\beta \over 2 \sqrt{L}} \sum_{\kappa=1}^{L-1}
     {1 \over \tilde{J}_\kappa} \tilde{h}_{-\kappa} \tilde{h}_\kappa \right\} \,.
 }
We are interested in the two-point function
 \eqn{GijDef}{
  G_{ij} = \langle \phi_i \phi_j \rangle = 
   {1 \over \beta^2 Z[0]} \left. {\partial^2 Z[h] \over \partial h_i \partial h_j} \right|_{h=0} \,.
 }
From $J_{ij} = J_{i-j}$ it follows that $G_{ij} = G_{i-j}$.  A short calculation starting with \eno{GijDef} leads to
 \eqn{GhGeneral}{
  G_h = -{1 \over \beta L^{3/2}} \sum_{\kappa=1}^{L-1} 
    {1 \over \tilde{J}_\kappa} e^{2\pi i \kappa h/L} \,.
 }
The factor of $\delta(\tilde\phi_0)$ in the partition function may seem undesirable, especially from the point of view of constructing Hamiltonians with only sparse couplings among the spins, because $\delta(\tilde\phi_0)$ can be viewed as the $K \to \infty$ limit of $e^{-K\tilde\phi_0^2}$, and this amounts to a strong all-to-all coupling among spins (though of a very particular form).  In fact, we could achieve the essentially the same results by omitting the factor of $\delta(\tilde\phi_0)$ while sending $J_0 \to J_0 - \mu$ where $\mu$ is small and positive.  Then $\tilde{J}_0 \propto -\mu$, while the other $\tilde{J}_\kappa$ would scarcely be affected since they are finite and negative already at ${\cal O}(\mu^0)$.  Use of \eno{GijDef} would then lead to the same $G_h$ as in \eno{GhGeneral}, up to an overall constant proportional to $1/\mu$.  Discarding this uninteresting constant and then taking the limit $\mu \to 0$ would lead to precisely the result given in \eno{GhGeneral}.  In other words, we can recover \eno{GhGeneral} by starting with a massive theory with truly sparse couplings and taking the massless limit.

In section \ref{SPARSE} we will provide the exact formulation of the sparse coupling model that is the main subject of this paper. But first we will apply the analysis leading to \eno{GhGeneral} in considering the Archimedean and $p$-adic statistical models that the sparse coupling model interpolates between as the spectral parameter ranges from negative to positive values.

\subsection{Nearest neighbor coupling}

As an extremal case of an Archimedean statistical model, we consider the model with nearest neighbor coupling specified by
 \eqn{JNN}{
  J_h^{\rm NN} = J_* (\delta_{h+1} + \delta_{h-1} - 2\delta_h) \,,
 }
which leads to
 \eqn{GhNN}{
  G^{\rm NN}_h = {1 \over 4 \beta J_* L} \sum_{\kappa=1}^{L-1} 
   {e^{2\pi i \kappa h/L} \over \sin^2 {\pi\kappa \over L}} \,.
 }
If $L$ is large, then we can approximate $\sin {\pi\kappa \over L} \approx {\pi\kappa \over L}$ and extend the sum to infinity:
 \eqn{GhNNapprox}{
  G^{\rm NN}_h \approx {L \over \beta J_*} \sum_{\kappa=-\infty,\,\, \kappa \neq 0}^\infty 
    {e^{2\pi i \kappa h/L} \over 4\pi^2 \kappa^2}
    = {L \over \beta J_*} G(h/L) \,,
 }
where the continuum two-point function $G(x)$ takes the form
 \eqn{Gx}{
  G(x) = {1 \over 2} \left( x - {1 \over 2} \right)^2 - {1 \over 24} \qquad\hbox{for}\qquad x \in [0,1] \,.
 }
Properly speaking, $G(x)$ is defined on a circle with $x \sim x+1$, with periodic boundary conditions, and it satisfies
 \eqn{GreenSatisfies}{
  {d^2 G \over dx^2} = -\delta(x) + 1 \qquad\hbox{and}\qquad
   \int_0^1 dx \, G(x) = 0 \,.
 }
If instead of nearest neighbor coupling we have some generic finite-range $J_h$ satisfying $J_h = J_{-h}>0$ for $h\neq 0$ and $\tilde{J}_0 = 0$, then we get essentially the same result: 
 \eqn{JkappaApprox}{
  \tilde{J}_\kappa \approx -{4\pi^2 \kappa^2 \over L^{5/2}} J_* \qquad\hbox{for}\qquad
   \left| {\kappa \over L} \right|_\infty \ll 1
 }
for some positive constant $J_*$, and so for large $L$,
 \eqn{Gh}{
  G_h \approx {L \over \beta J_*} G(h/L)
 }
with the same continuum function $G(x)$ given in \eno{Gx}.

It is worth noting that if we focus in on small $|h/L|_\infty$, then we are mostly insensitive to the fact that the system is at finite volume, and we find $G(x) \approx G(0) - |x|_\infty/2$.

\subsection{Power-law coupling}
\label{POWER}

For comparison with the sparse coupling model to be defined in section \ref{SPARSE}, we will eventually need to adjust the nearest neighbor model so as to have it include an adjustable exponent that tunes the strength of the coupling, analogously to the spectral parameter of the sparse coupling model. To this end we define
 \eqn{JpowerDef}{
  \tilde{J}^{\rm power}_\kappa \equiv -\frac{J_*}{2^s\sqrt{L}} \left[ \sin\left( 
    {\pi\kappa \over L} \right) \right]^{-s}
 }
so that
 \eqn{GpowerDef}{
  \tilde{G}^{\rm power}_\kappa = {2^s \over \beta J_* \sqrt{L}} \left[ \sin\left( 
    {\pi\kappa \over L} \right) \right]^s \,.
 }
 For $s=-2$, this model reduces to the nearest neighbor coupling model.  In general for $s<1$, one can approximate the Fourier series of \eqref{JpowerDef} with an integral in the limit $h/L \rightarrow 0$ to find that 
  \eqn{JpowerLargeL}{
  J^{\rm power}_h \sim -\frac{J_*}{\pi}\,\frac{\Gamma(1-s)\sin(\pi s/2)\Gamma(h+s/2)}{\Gamma(1+h-s/2)}.
 }
 By additionally invoking Sterling's formula, it becomes apparent that in the regime $1 \ll h \ll L$, the model we are considering does indeed couple the spins according to a power law:
   \eqn{JpowerLargeLagain}{
 J^{\rm power}_h \sim -\frac{J_*}{\pi}\,\Gamma(1-s)\sin\left(\pi s/2\right) h^{s-1}.
 }
 For $s<-1$, the large $L$ limit of the position space Green's function asymptotes to
\eqn{Ghpower}{
  G^{\rm power}_h = {2^s\pi^s \over \beta J_* L^{1+s}} \big[ \text{Li}_{-s}(e^{2\pi i h/L})+\text{Li}_{-s}(e^{-2\pi i h/L}) \big] \,,
  }
where Li$_{n}(x)$ denotes the polylogarithm function.

\subsection{\texorpdfstring{$p$-}adic coupling}

Choose a prime number $p$ and a positive integer $N$, and assume
 \eqn{Ldef}{
  L = p^N \,.
 }
Then an all-to-all coupling of spins can be defined based on the $p$-adic norm:
 \eqn{padicCoupling}{
  J^{p-\rm adic}_h = \left\{ \seqalign{\span\TL & \qquad\span\TT}{
   J_* |h|_p^{-s-1} & if $h \neq 0$ \cr
   -J_* L {\zeta_p(-s) \over \zeta_p(1) \zeta_p(-Ns)} & if $h = 0$\,.} \right.
 }
Here we have used the local zeta function
 \eqn{zetaDef}{
  \zeta_p(s) \equiv {1 \over 1-p^{-s}} \,,
 }
so named because the usual Riemann zeta function is $\zeta(s) = \prod_p \zeta_p(s)$ where the product is over all prime numbers.

To analyze \eno{padicCoupling}, it is useful first to work out the Fourier transform of the following function:
 \eqn{padicFunction}{
  f_h = A |h|_p^{-s-1} (1-\delta_h) + B + C \delta_h \,.
 }
A tedious but straightforward calculation suffices to show that
 \eqn{padicFourier}{
  \tilde{f}_\kappa = \tilde{A} |\kappa|^s (1-\delta_\kappa) + \tilde{B} + \tilde{C} \delta_\kappa
 }
where
 \eqn{FTA}{
  \tilde{A} = L^{s+{1 \over 2}} {\zeta_p(-s) \over \zeta_p(1+s)} A \qquad\quad
  \tilde{B} = {C \over \sqrt{L}} - L^{s+{1 \over 2}} {\zeta_p(-s) \over \zeta_p(1)} A \qquad\quad
  \tilde{C} = \sqrt{L} \left( B + {\zeta_p(-s) \over \zeta_p(1)} A \right) \,.
 }
With the help of \eno{FTA} one can see immediately that $J^{p-\rm adic}_0$ was chosen in \eno{padicCoupling} precisely so as to have $\tilde{J}^{p-\rm adic}_0 = 0$.  Indeed,
 \eqn{JpadicFourier}{
  \tilde{J}^{p-\rm adic}_\kappa = J_* \sqrt{L} \left[ {\zeta_p(-s) \over \zeta_p(1+s)} 
   \left| {\kappa \over L} \right|_p^s - 
    {\zeta_p(-s) \over \zeta_p(1)} \right] (1-\delta_\kappa) \,.
 }
While $\tilde{J}^{p-\rm adic}_\kappa < 0$ for $\kappa \neq 0$ for any $s \in \mathbb{R}$, we are mostly interested in the regime $s>0$, in which case the absolute value of the first term in square brackets in \eno{JpadicFourier} is larger than the absolute value of the second.  Thus we may expand
 \eqn{GpadicFourier}{
  \tilde{G}^{p-\rm adic}_\kappa = -{1 \over \beta L \tilde{J}_\kappa} (1-\delta_\kappa) = 
   -{\zeta_p(1)/\zeta_p(-s) \over \beta L^{3/2} J_*} \sum_{n=1}^\infty
     \left( {\zeta_p(1+s) \over \zeta_p(1)} L^{-s} \right)^n |\kappa|_p^{-ns} (1-\delta_\kappa) \,.
 }
The expansion is useful because it allows us to apply the Fourier transform \eno{padicFunction}-\eno{FTA} and obtain
 \eqn{Gpadic}{
  G^{p-\rm adic}_h &= -{\zeta_p(1)/\zeta_p(-s) \over \beta L^2 J_*} \sum_{n=1}^\infty 
    {\zeta_p(1+s)^n \over \zeta_p(1)^n}
   \Bigg[ \left( {\zeta_p(-ns+1) \over \zeta_p(ns)} |h|_p^{ns-1} - 
    {\zeta_p(-ns+1) \over \zeta_p(1)} \right) (1-\delta_h)  \cr
   &\qquad\qquad\qquad{} - {\zeta_p(-ns+1) \over \zeta_p(1) \zeta_p(N(ns-1))} \delta_h \Bigg] \,.
 }
In a sense, the result \eno{Gpadic} is more complicated than necessary, because by adding a constant to $J^{p-\rm adic}_h$ for $h \neq 0$ and adjusting $J^{p-\rm adic}_0$ to keep $\tilde{J}^{p-\rm adic}_0=0$, we could have arranged to have $\tilde{J}^{p-\rm adic}_\kappa = J_* \sqrt{L} {\zeta_p(-s) \over \zeta_p(1+s)} \left| {\kappa \over L} \right|_p^s$, which would result in the same result \eno{Gpadic} for $G^{p-\rm adic}_h$, except with the infinite sum replaced by its first term: That is, $G^{p-\rm adic}_h = A |h|_p^{s-1} + B + C \delta_h$ for some constants $A$, $B$, and $C$ depending on $s$ and proportional to ${1 \over \beta L^2 J_*}$.  However, for purposes of analyzing the next example, the alterations in $J^{p-\rm adic}_h$ just described are undesirable.

Note that if we hold $L^2 J_*$ fixed, then except at $h=0$ there is no $L$ dependence at all in $G^{p-\rm adic}_h$; the only thing that changes is the range of allowed $h$.  Taking $L$ large means that the range of $h$ becomes $p$-adically dense in the $p$-adic integers $\mathbb{Z}_p$, defined as the subset of $\mathbb{Q}_p$ consisting of elements whose norm is no greater than $1$.  $\mathbb{Z}_p$ can be understood as the $p$-adic analog of the interval $[-1,1] \subset \mathbb{R}$.  Because $G_h^{p-\rm adic}$ is a function of $h$ only through its $p$-adic norm $|h|_p$, we see that its continuum limit is locally constant everywhere on $\mathbb{Z}_p$, except at $h=0$.

The results of this section are perhaps not too surprising when compared with power-law interactions in real field theories.  Indeed, a power law $1/|x|_\infty^\alpha$ in the action leads to a power law $1/|x|_\infty^{\tilde\alpha}$ in the Green's functions, where $\tilde\alpha + \alpha = 2d$ and $d$ is the dimension of the field theory (c.f.~results in section~\ref{POWER}.)  The current setup is essentially the same, except that the ordinary absolute value has been replaced by the $p$-adic norm.

\subsection{The main model of interest: Sparse coupling}
\label{SPARSE}

Now let
 \eqn{Lspecial}{
  L = 2^N
 }
for some positive integer $N$.  Then we can consider a sparse coupling of the form
 \eqn{Jsparse}{
  J^{\rm sparse}_h = 
    J_* \sum_{n=0}^{N-1} 2^{ns} (\delta_{h-2^n} + \delta_{h+2^n} - 2\delta_h) \,.
 }
By sparse we mean that out of $L$ independent values of $J_h$, only ${\cal O}(\log L)$ are non-vanishing.  We could generalize from $p=2$ to other values of $p$, but some unobvious complications arise in doing so which we prefer to postpone.

The main qualitative features of $G^{\rm sparse}_h$ are:
 \begin{itemize}
  \item For sufficiently negative $s$, $G^{\rm sparse}_h$ closely approximates $G^{\rm NN}_h$.  This makes sense because when $s$ is large and negative, only the first few terms in the sum matter.
  \item For sufficiently positive $s$, $G^{\rm sparse}_h$ closely approximates $G^{2-\rm adic}_h$.  This is less obvious and will be investigated further in the next section.
  \item As $s$ crosses from negative to positive values, $G^{\rm sparse}_h$ undergoes a transition from being closer to a smooth function in an Archimedean sense to being closer to a smooth function in a $2$-adic sense.
 \end{itemize}

To visualize the behavior of $G^{\rm sparse}_h$, we have found it helpful to employ a discrete version of the Monna map, introduced for $p=2$ already in figure~\ref{FurthestNeighbor}.  For completeness, we record here its definition for any $p$.  Let any $h \in \{0,1,2,\dots,L-1\}$ be expressed as
 \eqn{basep}{
  h = \sum_{n=0}^{N-1} h_n p^n \qquad\hbox{where each $h_n \in \{ 0,1,2,\dots,p-1 \}$ \,.}
 }
Then the image of $h$ under the Monna map is
 \eqn{Monna}{
  \mathfrak{M}(h) \equiv \sum_{n=0}^{N-1} h_{N-1-n} p^n \,.
 }
That is, we reverse the digits in the base $p$ expansion of $h$.  Clearly (this version of) the Monna map is an involution.\footnote{The standard Monna map from $\mathbb{Q}_p$ to the non-negative reals is defined similarly, by expanding $x \in \mathbb{Q}_p$ as $x = \sum_{n=v(x)}^\infty x_n p^n$ and then defining $\mathfrak{M}(x) \equiv \sum_{n=v(x)}^\infty x_n p^{-1-n}$.  This map is continuous, volume-preserving, and surjective, but not quite injective: For example, $\mathfrak{M}$ maps both $-1$ and $1/p$ to $1$.  We will not have need of this continuous version of the Monna map.}  By inspection, we see that if $i$ and $j$ are $p$-adically close, then $\mathfrak{M}(i)$ and $\mathfrak{M}(j)$ are sequentially close.

In figure~\ref{ComparisonPlots} we show $G^{\rm sparse}_h$ and $G^{2-\rm adic}_h$, the former as a function of both $h$ and $\log_2 \mathfrak{M}(h)$, for various values of $s$, to confirm the qualitative features listed above.
 \CaptionedFigure[6in]{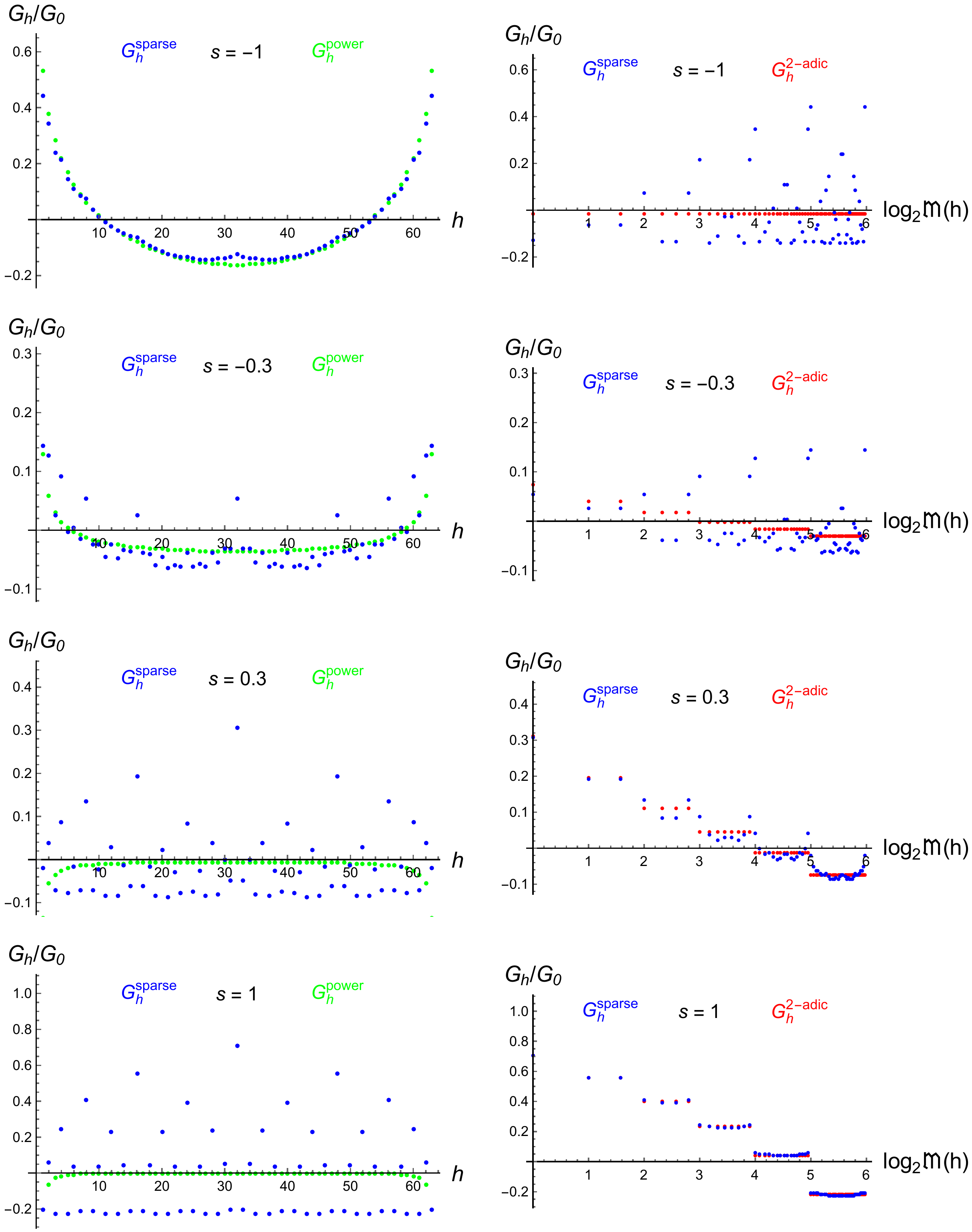}{
 Left: $G^{\rm sparse}_h$ and $G^{\rm power}_h$ versus $h$. This column shows how close $G^{\rm sparse}_h$ is to a smooth function in the usual Archimedean sense, and confirms that $G^{\rm sparse}_h \approx G^{\rm power}_h$ when $s$ is sufficiently negative.
 \\
 Right: $G^{\rm sparse}_h$ and $G^{2-\rm adic}_h$ versus $\log_2 \mathfrak{M}(h)$. This column shows how close $G^{\rm sparse}_h$ is to a smooth function in the $2$-adic sense, and confirms that $G^{\rm sparse}_h \approx G^{2-\rm adic}_h$ when $s$ is sufficiently positive.}
\clearpage

\section{Continuity bounds}
\label{CONTINUITY}

Having observed an apparent change from Archimedean to $2$-adic continuity in the example of section~\ref{SPARSE}, we are naturally led to investigate continuum theories with similar coupling patterns.  We start in section~\ref{PADIC} with $p$-adic field theories, since they are actually easier to deal with once one understands the rules than Archimedean field theories.  We derive H\"older continuity bounds for the two-point Green's function both in momentum space and real space.  Then in section~\ref{ARCHIMEDEAN} we derive analogous bounds for bilocal Archimedean field theories.

Before getting into the main field theory calculations, let's review what H\"older continuity bounds are in general.  Let $F$ be either $\mathbb{Q}_p$ or $\mathbb{R}$, and denote the norm on $F$ as $|\cdot|$.  Let $f$ be a map from some subset $D \subset F$ to $\mathbb{R}$.  Usually, if $F = \mathbb{R}$, then for us $D$ will be an open interval, while if $F = \mathbb{Q}_p$, then $D$ will be an affine copy of $\mathbb{Z}_p$.  Let $O$ be any subset of $D$ (and again we usually have in mind simple choices of $O$ like open intervals or affine copies of $\mathbb{Z}_p$).  Then $f$ satisfies a H\"older continuity condition over $O$ with positive real exponent $\alpha$ iff there is some positive real number $K$ such that
 \eqn{HolderDef}{
  |f(x_1)-f(x_2)|_\infty < K |x_1-x_2|^\alpha
 }
for all $x_1$ and $x_2$ in $O$.  If $O=D$, then we would say that $f$ is globally $\alpha$-H\"older continuous.  We say that $f$ is locally $\alpha$-H\"older continuous at $x$ iff there exists some open set $I$ containing $x$ such that $f$ is $\alpha$-H\"older continuous on $I$.  And we describe $f$ as a whole as locally $\alpha$-H\"older continuous if it is locally $\alpha$-H\"older continuous at every point in its domain (assumed to be an open set).

A H\"older continuous function with any positive exponent $\alpha$ is continuous in the usual sense.  How big we can make $\alpha$ is an indication of how much ``better'' our function is than merely continuous.  If $F = \mathbb{R}$, then we don't usually expect to find $\alpha$ bigger than $1$, because if we do, then $f$ must be constant over its connected components.  But if $F = \mathbb{Q}_p$, then it is possible to have non-constant functions with arbitrarily positive H\"older continuity exponent.  A useful example of an $\alpha$-H\"older continuous function $f(x)$ is a linear combination of functions $|x-x_i|^\alpha$ where the $x_i$ are constants.

The distinction between global and local $\alpha$-H\"older continuity is important to us because we are going to argue, through a combination of analytic and numerical means, that the continuum limit of the two-point function $G^{\rm sparse}_h$ is, in some cases, globally H\"older continuous with one exponent and locally H\"older continuous away from the origin with a larger exponent.

\subsection{\texorpdfstring{$2$-}{2-}adic field theories}
\label{PADIC}

The standard integration measure on $\mathbb{Q}_p$ satisfies two key properties:
 \begin{itemize}
  \item The measure of $\mathbb{Z}_p$ is $1$.
  \item If $S \subset \mathbb{Q}_p$ has measure $\ell$ (a real number), then the set $aS + b$ has measure $|a|_p\ell$ for any $a,b \in \mathbb{Q}_p$.
 \end{itemize}
The Fourier transform on $\mathbb{Q}_p$ is defined by
 \eqn{FourierQp}{
  f(x) = \int_{\mathbb{Q}_p} dk \, \chi(kx) \tilde{f}(k)
 }
where $\chi(kx) = e^{2\pi i \{kx\}}$.  The notation $\{\xi\}$ means the fractional part of $\xi \in \mathbb{Q}_p$: that is, $\{\xi\} = \xi + n$ for the unique element $n \in \mathbb{Z}_p$ that leads to $\{\xi\} \in [0,1)$.  Just as with ordinary plane waves on $\mathbb{R}$, we have $\chi(\xi_1+\xi_2) = \chi(\xi_1)\chi(\xi_2)$; in technical terms, $\chi$ is an additive character.  Note that $\chi(\xi) = 1$ precisely if $\xi \in \mathbb{Z}_p$.

Specializing now to $p=2$, consider the bilocal field theory
 \eqn{SJQp}{
  S = -\int_{\mathbb{Q}_2} dx dy \, {1 \over 2} \phi(x) J(x-y) \phi(y)
 }
where
 \eqn{JsparseDelta}{
  J(x) = J_* \sum_{n \in \mathbb{Z}} 2^{ns} \left[ \delta(x-2^n) + \delta(x+2^n) -
    2 \delta(x) \right] \,,
 }
and $\delta(x)$ is defined as usual by the relation $\int_{\mathbb{Q}_2} dx \, f(x) \delta(x) = f(0)$ for any continuous function $f$.  The action \eno{SJQp} becomes diagonal in Fourier space:
 \eqn{SJQpFourier}{
  S = -\int_{\mathbb{Q}_2} dk \, {1 \over 2} \tilde\phi(-k) \tilde{J}(k) \tilde\phi(k) \,,
 }
The two-point function is defined as
 \eqn{TwoPoint}{
  G(x) = \langle \phi(x) \phi(0) \rangle \equiv 
   {\int {\cal D}\phi \, e^{-S} \phi(x) \phi(0) \over \int {\cal D}\phi \, e^{-S}} \,,
 }
and one straightforwardly finds
 \eqn{Gk}{
  \tilde{G}(k) = -{1 \over \tilde{J}(k)} \,.
 }
For explicit calculations, it is convenient to set $J_* = 1/4$.  Then
 \eqn{Jk}{
  \tilde{J}(k) = {1 \over 4} 
    \sum_{n \in \mathbb{Z}} 2^{ns} \left[ \chi(2^n k) + \chi(-2^n k) - 2
   \right] = 
  -\sum_{n \in \mathbb{Z}} 2^{ns} \sin^2(\pi \{ 2^n k \}) \,.
 }
The infinite sums in \eno{Jk} may be restricted to $n < -v(k)$, because only then is $\{2^n k\}$ non-zero.\footnote{An amusing connection to population dynamics can be observed at this point.  Recall the logistical map, $x \to r x(1-x)$.  If $\{x_n\}_{n\in\mathbb{Z}}$ is a solution to this iterated map, then we can think of $x_n$ as (proportional to) the population of a species at generation number $n$.  For $r=4$, a solution is $x_n = \sin^2(\pi 2^n k)$ where $k$ is a real number.  However, this is not the most general solution, because it has the property $x_n \to 0$ as $n \to -\infty$.  Consider instead $x_n = \sin^2(\pi \{ 2^n k \})$ where $k$ is a $2$-adic number.  Then $x_n=0$ for all $n \geq -v(k)$, but we need not have $x_n \to 0$ as $n \to -\infty$.  Thus we see that the $2$-adic number $k$ parametrizes the routes to extinction under the $r=4$ logistical map, and the $2$-adic norm of $k$ predicts the moment of extinction: $n_* = \log_2 |k|_2$.  To make the discussion simple, suppose now that $k$ is a $2$-adic integer, so that extinction has occurred by the time $n=0$.  Further suppose that each generation leaves an imprint on its environment proportional to $x_n$, and that this imprint dissipates over time with a half life of $1/s$ generations.  So the environmental imprint at time $0$ of generation $n$ (with $n<0$ since extinction occurs no later than time $0$) is $I_n = \alpha 2^{ns} x_n$, where $\alpha$ is the constant of proportionality.  Then $I = -\alpha\tilde{J}(k)$ as computed in \eno{Jk} is the total environmental imprint of the species, summed across all generations and measured at time $0$.}  We immediately see that it is necessary to choose $s>0$ in order to have convergence.

Assuming $s>0$, we may rewrite \eno{Jk} for non-zero $k$ as
 \eqn{JPsi}{
  \tilde{J}(k) = -|k|_2^s \Psi(\hat{k})
 }
where $\hat{k} = |k|_2 k$ and
 \eqn{PsiDef}{
  \Psi(\hat{k}) &\equiv \sum_{n=1}^\infty 2^{-ns} \sin^2(\pi \{ 2^{-n} \hat{k} \}) \,.
 }
The following features of $\Psi(\hat{k})$ are at the center of our analysis:
 \begin{enumerate}
  \item $\Psi(\hat{k})$ is bounded above and below by positive constants which depend on $s$ but not $\hat{k}$.
  \item $\Psi(\hat{k})$ is globally $s$-H\"older continuous over $\mathbb{U}_2$.
 \end{enumerate}
The first of these properties is easily demonstrated:
 \eqn{PsiBounded}{
  2^{-s} + 2^{-2s-1} \leq \Psi(\hat{k}) \leq -\zeta_2(-s) \,,
 }
where the first inequality comes from dropping all but the first two terms in the sum \eno{PsiDef}, and the second inequality comes from replacing $\sin^2(\pi \{2^{-n} \hat{k}\})$ by $1$ in all terms of the sum.

The second property requires more care, and it turns on observing that if $n \leq v(\hat{k}_1 - \hat{k}_2)$, then $\sin^2(\pi \{ 2^{-n} \hat{k}_1 \}) = \sin^2(\pi \{ 2^{-n} \hat{k}_2 \})$.  (This follows because if $n \leq v(\hat{k}_1 - \hat{k}_2)$, then $2^{-n} \hat{k}_1$ and $2^{-n} \hat{k}_2$ differ by a $2$-adic integer, so $\chi(2^{-n} \hat{k}_1) = \chi(2^{-n} \hat{k}_2)$.)  Therefore, when computing $\Psi(\hat{k}_1) - \Psi(\hat{k}_2)$, only the terms with $n > v(\hat{k}_1 - \hat{k}_2)$ contribute, and if we replace $\sin^2(\pi \{2^{-n} \hat{k}\})$ by $1$ in these terms we arrive at the desired H\"older inequality with
 \eqn{Kvalue}{
  K = -\zeta_2(-s) \,.
 }

The boundedness property of $\Psi(\hat{k})$ implies that $1/\Psi(\hat{k})$ is also globally $s$-H\"older continuous.  Since the Green's function
 \eqn{GPsi}{
  \tilde{G}(k) = {1 \over |k|_2^s \Psi(\hat{k})} \,,
 }
is the product of a locally constant factor and a H\"older-continuous factor, we conclude that away from $k=0$, $\tilde{G}(k)$ is locally $s$-H\"older continuous.

Turning to position space, our intuitive understanding is that $G(x)$ will be continuous everywhere iff $\tilde{G}(k)$ is integrable at large $k$, which is the case iff $s>1$.  Let us focus therefore on the regime $s>1$.  There is a complication in defining $G(x)$ when $s>1$: The integral
 \eqn{GxResult}{
  G(x) = \int_{\mathbb{Q}_2} dk \, {\chi(kx) \over |k|_2^s \Psi(\hat{k})}
 }
is infrared divergent.  An efficient way to handle this divergence is to alter \eno{GxResult} to
 \eqn{GxReg}{
  G(x) \equiv \int_{\mathbb{Q}_2} dk \, {\chi(kx)-1 \over |k|_2^s \Psi(\hat{k})} = 
    |x|_2^{s-1} g(\hat{x}) \,,
 }
where, by calculation,
 \eqn{gxDef}{
  g(\hat{x}) = \zeta_2(1-s) \int_{\mathbb{U}_2} {d\hat{k} \over \Psi(\hat{k})} + 
    \sum_{n=1}^\infty 2^{(1-s)n} \int_{\mathbb{U}_2} d\hat{k} \,
     {\chi(2^{-n} \hat{k} \hat{x}) \over \Psi(\hat{k})} \,.
 }
Other approaches to regulating the infrared divergence give substantially the same result.\footnote{For example, instead of \eno{GxReg} we could stick with \eno{GxResult} but exclude from the domain of integration all $k$ with $|k|_2 < |k_{\rm IR}|_2$, where $k_{\rm IR} = 2^{v_{\rm IR}}$ is an infrared regulator (with $v_{\rm IR}$ large and positive).  Then we would find
 \eqn{GxRegIR}{
  G(x) = \zeta_2(s-1) |k_{\rm IR}|^{1-s} \int_{\mathbb{U}_2} {d\hat{k} \over \Psi(\hat{k})} + 
     |x|_2^{s-1} g(\hat{x}) \,,
 }
and upon dropping the first term we are back to \eno{GxReg}.}

We can conclude from \eno{GxReg} that $G(x)$ is globally H\"older continuous with exponent $s-1$, provided we can show $g(\hat{x})$ is globally H\"older continuous with the same exponent.  (Note that the H\"older bound for $G(x)$ can be made global rather than local because $|x|_2^{s-1}$ is itself globally H\"older continuous with exponent $s-1$.)  To that end, we note that when computing the difference $g(\hat{x}_1) - g(\hat{x}_2)$, we can restrict the sum in \eno{gxDef} to $n > v(\hat{x}_1 - \hat{x}_2)$.  The remaining terms can be bounded using $|\chi(2^{-n}\hat{k}\hat{x}_1) - \chi(2^{-n}\hat{k}\hat{x}_2)|_\infty \leq 2$, and the desired H\"older condition follows.  Note that our final position space continuity condition is significantly weaker than the one in momentum space, because the H\"older exponent, which was $s$ in momentum space, is now $s-1$.  In section~\ref{NUMERICS}, we will in fact find numerical evidence that a stronger H\"older condition is possible {\it locally} in position space, away from $x=0$.  No improvement to the {\it global} H\"older continuity exponent is possible, though, because if it were we could demonstrate a faster fall-off of $\tilde{G}(k)$ at large $|k|_2$ than the one that follows from \eno{GPsi}.

\subsection{Archimedean field theory}
\label{ARCHIMEDEAN}

A similar analysis can be carried out on the Archimedean side, starting with the field theory
 \eqn{SJR}{
  S = -\int_{\mathbb{R}} dx dy \, {1 \over 2} \phi(x) J(x-y) \phi(y) = 
    -\int_{\mathbb{R}} dk \, {1 \over 2} \tilde\phi(-k) \tilde{J}(k) \tilde\phi(k) \,,
 }
where the Fourier transform is
 \eqn{FourierPhi}{
  f(x) = \int_{\mathbb{R}} dk \, e^{2\pi i k x} \tilde{f}(k) \,,
 }
and we use precisely the same form of $J(x)$ as in \eno{JsparseDelta}.  The general analysis \eno{TwoPoint}-\eno{Gk} of two-point functions holds unaltered, now leading to
 \eqn{JkReal}{
  \tilde{J}(k) = -\psi(k)
 }
where we have set $J_* = 1/4$ for convenience, and
 \eqn{psiDef}{
  \psi(k) \equiv \sum_{n\in \mathbb{Z}} 2^{ns} \sin^2(\pi 2^n k) \,.
 }
If $s \leq -2$, the infinite sum in \eno{psiDef} diverges at large negative $n$.\footnote{Again a population dynamical narrative can be attached to (a slight variant of) \eno{psiDef}: Regarding $x_n = \sin^2 (\pi 2^n k)$ as a solution to the $r=4$ logistical map, and supposing that each generation ``eats'' an amount $\alpha x_n$ of a resource which, when undisturbed, grows exponentially with doubling time $-1/s$, we see that a cutoff version of the sum, $\psi_>(k) \equiv \sum_{n=0}^\infty 2^{ns} \sin^2(\pi 2^n k)$, computes for us the total quantity of resources $I = \alpha \psi_>(k)$ required at time $0$ to feed the species for all future time.  Here $\alpha$ is some constant of proportionality.}  But this only means that the coupling function $J(x)$ is overwhelming concentrated near $x=0$.  If a cutoff is imposed on the sum, and then $J_*$ is rescaled as this cutoff is gradually removed, one can show that $J(x)$ converges precisely to $-\delta''(x)$, resulting in a perfectly local theory.  If instead $s \geq 0$, then the sum in \eno{JkReal} diverges at large positive $n$, signaling that arbitrarily long-ranged interactions dominate.

The interesting regime, then, is $-2 < s < 0$.  Here the sum \eno{psiDef} converges, and we can ask what properties the function $\psi(k)$ satisfies analogous to the ones enumerated below \eno{PsiDef} for $\Psi(\hat{k})$ in the $2$-adic case.  In fact, we claim
 \begin{enumerate}
  \item $\psi(k) \approx |k|_\infty^{-s}$, meaning that there exist positive constants $K_1$ and $K_2$, independent of $k$, such that $K_1 |k|_\infty^{-s} < \psi(k) < K_2 |k|_\infty^{-s}$ for all $k \in \mathbb{R} \backslash \{0\}$.
  \item For $-1<s<0$, $\psi(k)$ is globally H\"older continuous with exponent $-s$.
  \item For $-2<s<-1$, the derivative $\psi'(k) = d\psi(k)/dk$ is globally H\"older continuous with exponent $-s-1$.  (Note that $\psi(k)$ itself cannot have H\"older continuity exponent greater than $1$ without being constant.  So the derivative condition we claim here is the best that can be expected.)  It follows that $\psi(k)$ is globally $1$-H\"older continuous on any bounded domain.
 \end{enumerate}

To arrive at the estimate $\psi(k) \approx |k|_\infty^{-s}$, we define
 \eqn{nkDef}{
  n_k \equiv -\log_2(\pi |k|_\infty) \,.
 }
Then we have
 \eqn{psiEst}{
  \psi(k) &= \sum_{n<n_k} 2^{ns} \sin^2(\pi 2^n k) + 
    \sum_{n \geq n_k} 2^{ns} \sin^2(\pi 2^n k)
   \approx \sum_{n<n_k} 2^{ns} (\pi 2^n k)^2 + \sum_{n \geq n_k} 2^{ns}  \cr
   &\approx (\pi |k|_\infty)^{-s} \left[ -\zeta_2(-s-2) + \zeta_2(-s) \right] \,.
 }
To derive the H\"older condition on $\psi$ for $-1<s<0$, set $\delta = |k_1-k_2|_\infty$ and note that
 \eqn{sinBound}{
  |\sin^2(\pi 2^n k_1) - \sin^2(\pi 2^n k_2)|_\infty \leq \min\{ 1, \pi 2^n \delta \} \,.
 }
Defining
 \eqn{ndeltaDef}{
  n_\delta = -\log_2(\pi\delta) \,,
 }
we see that
 \eqn{psiDiffEst}{
  |\psi(k_1) - \psi(k_2)|_\infty &\leq \sum_{n \in \mathbb{Z}} 2^{ns} \min\{ 1,\pi 2^n \delta \}
    = \sum_{n < n_\delta} 2^{n(s+1)} \pi\delta + \sum_{n \geq n_\delta} 2^{ns}  \cr
    &\approx (\pi\delta)^{-s} \left[ -\zeta_2(-s-1) + \zeta_2(-s) \right] \,,
 }
where again $\approx$ means equality to within fixed multiplicative factors, independent in this case of $\delta$.  The last expression in \eno{psiDiffEst} is the desired H\"older bound, valid when $-1<s<0$.  If instead $-2<s<-1$, then we may calculate
 \eqn{JkPrime}{
  \psi'(k) = \pi \sum_{n \in \mathbb{Z}} 2^{n(s+1)} \sin(\pi 2^{n+1}k) \,.
 }
By the same method as in \eno{psiDiffEst} we arrive at the H\"older continuity condition for $\psi'(k)$ with exponent $-s-1$.

By combining the property $\psi(k) \approx |k|_\infty^{-s}$ with the H\"older bounds, we see that $\tilde{G}(k)$ is locally H\"older with exponent $-s$ for $-1<s<0$.  Also, $\tilde{G}'(k)$ is locally H\"older away from $k=0$ with exponent $-s-1$ for $-2<s<-1$, implying that $\tilde{G}(k)$ is locally $1$-H\"older away from $k=0$.

Now let's investigate smoothness of the Green's function in position space.  We naively define
 \eqn{GxReal}{
  G(x) = -\int_{\mathbb{R}} dk \, {e^{2\pi i k x} \over \tilde{J}(k)} = 
    \int_{\mathbb{R}} dk \, {e^{2\pi i k x} \over \psi(k)} \,.
 }
As in the $2$-adic case, our intuitive understanding is that $G(x)$ will be continuous everywhere iff $\tilde{G}(k)$ is integrable at large $k$, which is the case iff $s < -1$.  Because $\psi(k) \approx |k|_\infty^s$, the UV-integrable regime is $-2<s<-1$ (where the lower limit is forced on us by the considerations explained following \eno{psiDef}), and in this regime the integral \eno{GxReal} has an IR divergence.  Again as in the $2$-adic case, the infrared divergence results in an overall additive constant in $G$.  It does not matter much how this constant is removed; one option is to alter \eno{GxReal} to
 \eqn{GxRegReal}{
  G(x) \equiv \int_{\mathbb{R}} dk {e^{2\pi i k x} - 1 \over \psi(k)}  \,.
 }
For the purposes of a H\"older continuity condition we must estimate
 \eqn{GxDiff}{
  G(x_1) - G(x_2) = \int_{\mathbb{R}} {dk \over \psi(k)} (e^{2\pi i k x_1} - 
    e^{2\pi i k x_2}) \,.
 }
Setting $\delta = |x_1-x_2|_\infty$, we have
 \eqn{GxEstReal}{
  |G(x_1) - G(x_2)|_\infty &\leq \int_{\mathbb{R}} {dk \over \psi(k)} 
    \min\{ 2, 2\pi |k|_\infty \delta \}
   = 2\pi \int_{|k| < 1/\delta} {dk \over \psi(k)} \, |k|_\infty \delta + 
     2 \int_{|k| > 1/\delta} {dk \over \psi(k)}  \cr
   &\approx \int_0^{1/\delta} dk \, k^{s+1} \delta + 
     \int_{1/\delta}^\infty k^s \approx \delta^{-s-1} 
       \left[ {1 \over s+2} - {1 \over s+1} \right] \,.
 }
In short, for $-2<s<-1$, we have obtained a global H\"older bound with exponent $-s-1$.

\section{Numerical evidence}
\label{NUMERICS}

\subsection{2-adic approximation of sparse coupling results}
\label{APPROXIMATION}

The first question we wish to ask of numerics is how well the two-point Green's function derived from sparse coupling approximates the one derived from $2$-adic coupling with the same value of $s$.  Based on the rigorous field theory results of section~\ref{CONTINUITY}, we expect that, for $s>0$, the answer in momentum space is that
 \eqn{Kineq}{
  K_1 < \tilde{G}^{\rm sparse}_\kappa / \tilde{G}^{2-\rm adic}_\kappa < K_2
 }
for some positive constants $K_1$ and $K_2$ which may depend on $s$.  Numerical support for this conclusion is shown in figure~\ref{SparseVsPadic}, where we show optimal values of $K_1$ and $K_2$ as functions of $s$ for various $N$.  As $s \to 0$, the evidence that $K_2$ remains bounded as $N$ increases becomes tenuous.  We are limited ultimately by our ability to go to sufficiently high values of $N$.  
 \CaptionedFigure[6.5in]{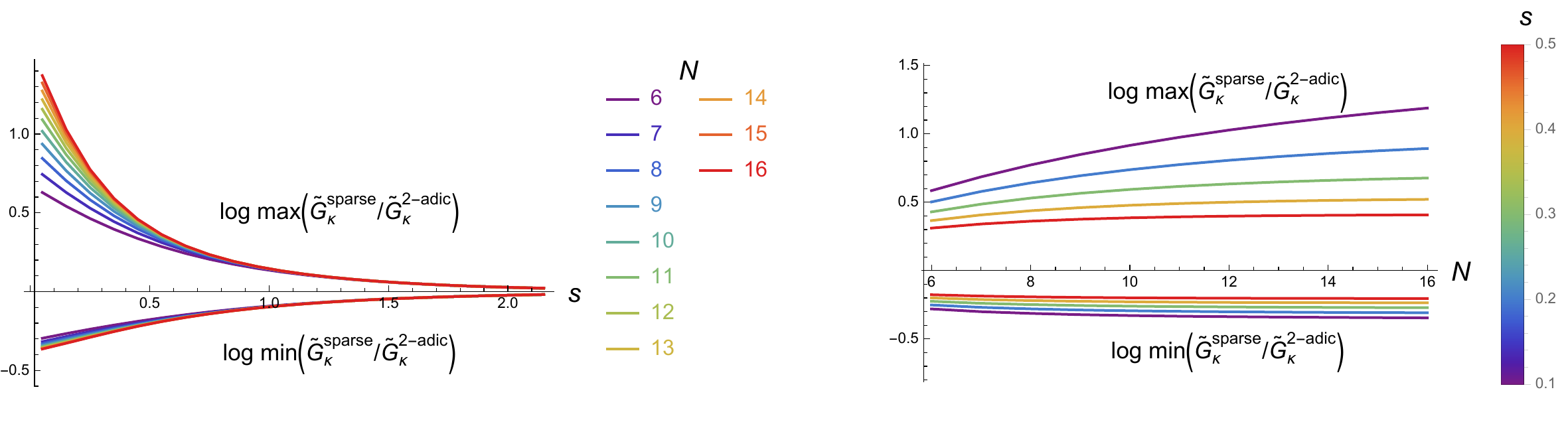}{Left: Optimal values of the constants $K_1$ and $K_2$ appearing in \eno{Kineq} as functions of $s$ for fixed $N$.\\
Right: Optimal values of the constants $K_1$ and $K_2$ as function of $N$ for fixed $s$.  The expectation is that provided $s>0$, $K_1$ and $K_2$ asymptote to constants at sufficiently large $N$.}
Away from small positive $s$, $\tilde{G}^{\rm sparse}_\kappa \approx \tilde{G}^{2-\rm adic}_\kappa$ is evidently an excellent approximation.  Based on empirically examining the curves on the left side of figure~\ref{SparseVsPadic}, we find $K_i \approx 1 + 2^{-2s} \varkappa_i(s)$ where the functions $\varkappa_i(s)$ vary relatively slowly with $s$, possibly as a negative power of $s$, or possibly as a small positive power of $2^{-s}$.  In order to obtain $K_1$ and $K_2$ as functions of $N$ and $s$, the actual procedure was as follows:
 \begin{enumerate}
  \item For fixed $N$ and $s$, compute $\tilde{G}^{\rm sparse}_\kappa$ using the methods of section~\ref{SETUP}, and adjust the overall coupling strength $J_*$ so that $G^{\rm sparse}_h = 1$ when $h=0$.  (In other words, the normalization condition is implemented in {\it position space}.)
  \item Likewise compute $\tilde{G}^{2-\rm adic}_\kappa$ with $G^{2-\rm adic}_0 = 1$.
  \item Compute $K_1$ and $K_2$ as
 \eqn{KonetwoDef}{
  K_1 &= \min\left( {\tilde{G}^{\rm sparse}_\kappa \over \tilde{G}^{2-\rm adic}_\kappa}
    \right) \equiv \min_{\kappa \neq 0}
     {\tilde{G}^{\rm sparse}_\kappa \over \tilde{G}^{2-\rm adic}_\kappa}  \cr
  K_2 &= \max\left( {\tilde{G}^{\rm sparse}_\kappa \over \tilde{G}^{2-\rm adic}_\kappa}
    \right) \equiv \max_{\kappa \neq 0}
     {\tilde{G}^{\rm sparse}_\kappa \over \tilde{G}^{2-\rm adic}_\kappa} \,.
 }
 \end{enumerate}

\subsection{Smoothness in momentum space}
\label{SmoothnessMomentum}

Next we would like to understand how well the local H\"older continuity bounds in momentum space are reflected in the numerics.  We also want to quantify how ragged the Green's functions become in momentum space in regimes where we couldn't derive any continuity bound (by methods developed in the current work).  The H\"older bounds, as derived in field theory in section~\ref{CONTINUITY}, are approximately as follows:
 \begin{itemize}
  \item $|\tilde{G}^{\rm sparse}(k_1)-\tilde{G}^{\rm sparse}(k_2)|_\infty < K |k_1-k_2|_2^s$ when $s>0$.  More precisely, $\tilde{G}^{\rm sparse}(k)$ as a map from $\mathbb{Q}_2$ to $\mathbb{R}$ is locally $s$-H\"older continuous away from $k=0$.
  \item $|\tilde{G}^{\rm sparse}(k_1)-\tilde{G}^{\rm sparse}(k_2)|_\infty < K |k_1-k_2|_\infty^{-s}$ when $-1<s<0$.  More precisely, $\tilde{G}^{\rm sparse}(k)$ as a map from $\mathbb{R}$ to $\mathbb{R}$ is locally $-s$-H\"older continuous away from $k=0$ when $-1<s<0$, and locally $1$-H\"older continuous away from $k=0$ when $-2<s<-1$.
 \end{itemize}

To test the H\"older bound on the $p$-adic side, we first adjust the overall coupling strength $J_*$ so that $G^{\rm sparse}_0=1$, and likewise $G^{2-\rm adic}_0=1$.  Then we define
 \eqn{Adef}{
  \tilde{A}^{2-\rm adic}(N,s) \equiv \log_2 \max_{\kappa\ \rm odd} 
     \left| {\tilde{G}^{\rm sparse}_\kappa \over \tilde{G}^{2-\rm adic}_\kappa} - 
       {\tilde{G}^{\rm sparse}_{\kappa+L/2} \over \tilde{G}^{2-\rm adic}_{\kappa+L/2}} 
     \right|_\infty \,,
 }
where on the right hand side we understand that $\tilde{G}^{\rm sparse}_\kappa$ and $\tilde{G}^{2-\rm adic}_\kappa$ are computed using the same values of $N$ and $s$.  We find numerically that $A^{2-\rm adic}(N,s)$ exhibits linear trajectories:
 \eqn{DeltaLaw}{
  \tilde{A}^{2-\rm adic}(N,s) \approx -s(N-1) + \log_2 \tilde{K}^{2-\rm adic}(s) \,,
 }
where $\tilde{K}^{2-\rm adic}(s)$ is $N$-independent.  These linear trajectories persist even at negative $s$, after $2$-adic continuity is lost.  See figure~\ref{kspaceSmoothness}.
 \CaptionedFigure[6.5in]{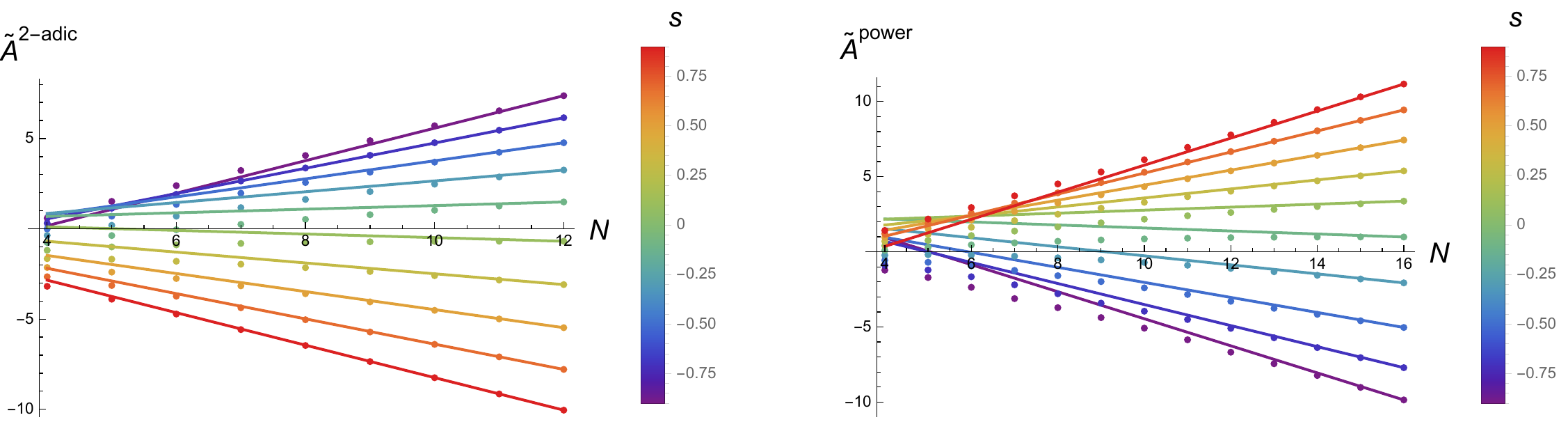}{Left: $2$-adic smoothness in momentum space.  The dots are evaluations of $\tilde{A}^{2-\rm adic}(N,s)$ in \eno{Adef}, and the lines are plots of the linear trajectories indicated in \eno{DeltaLaw}, with $K(s)$ chosen so that the line goes through the last data point.\\
Right: Archimedean smoothness in momentum space.  The dots are evaluations of $\tilde{A}^{\rm power}(N,s)$ in \eno{archDeltaDef}, and the lines are plots of the linear trajectories indicated in \eno{archDeltaLaw}, with $K(s)$ chosen so that the line goes through the last data point.}

In formulating the definition of $\tilde{A}^{2-\rm adic}(N,s)$, we chose to focus on differences between site $\kappa$ and $\kappa+L/2$ because these are nearest neighbors in terms of their $2$-adic norm.  To make the connection to H\"older continuity bounds more transparent, we note that \eno{Adef}-\eno{DeltaLaw} are equivalent to
 \eqn{HolderRephrase}{
  \left| {\tilde{G}^{\rm sparse}_{\kappa_1} \over \tilde{G}^{2-\rm adic}_{\kappa_1}} - 
   {\tilde{G}^{\rm sparse}_{\kappa_2} \over \tilde{G}^{2-\rm adic}_{\kappa_2}} \right|_\infty
    \leq 2^{\tilde{A}^{2-\rm adic}(N,s)} \approx
   \tilde{K}^{2-\rm adic}(s) \left| \kappa_1-\kappa_2 \right|_2^s
 }
for all odd $\kappa_1$ and $\kappa_2$ with $\kappa_2-\kappa_1 = L/2$.  The inequality \eno{HolderRephrase} is clearly a close relative of the local $s$-H\"older continuity condition on $\tilde{G}(k)$.  We could make an even closer connection to this continuity condition if we generalized $\tilde{A}^{2-\rm adic}(N,s)$ to a quantity that would track also the separation between $\kappa_1$ and $\kappa_2$.  Doing so would allow us to check the H\"older condition on $\tilde{G}^{\rm sparse}_\kappa / \tilde{G}^{2-\rm adic}_\kappa$ more thoroughly; however, our explorations in this direction seem to indicate that the final results are unaffected by such a generalization.

In light of the approximately linear trajectories \eno{DeltaLaw}, it is convenient to define
 \eqn{alphatDef}{
  \tilde\alpha^{2-\rm adic}(N,s) \equiv -\tilde{A}^{2-\rm adic}(N,s) + 
    \tilde{A}^{2-\rm adic}(N-1,s) \,.
 }
Then, recalling that $\tilde{G}^{2-\rm adic}_\kappa$ is a $2$-adically smooth function, we arrive at our main numerical result on $2$-adic smoothness of momentum space Green's functions: $\tilde{G}^{\rm sparse}_\kappa$ satisfies a local H\"older condition whose best (i.e.~most positive) exponent is approximately $\tilde\alpha^{2-\rm adic}(N,s) \approx s$, in agreement with our field theory expectations.

On the Archimedean side, in order to pursue a similar strategy, we need some standard of comparison analogous to $\tilde{G}^{2-\rm adic}_\kappa$.  We define
 \eqn{archDeltaDef}{
  \tilde{A}^{\rm power}(N,s) \equiv \log_2 \max_{{L \over 4} \leq \kappa < {3L \over 4}}
    \left| {\tilde{G}^{\rm sparse}_\kappa \over \tilde{G}^{\rm power}_\kappa} - 
       {\tilde{G}^{\rm sparse}_{\kappa+1} \over \tilde{G}^{\rm power}_{\kappa+1}} 
     \right|_\infty \,,
 }
 where $\tilde{G}^{\rm power}_\kappa$ is given in \eno{GpowerDef} as usual we can adjust $J_*$ so that $G^{\rm power}_h=1$ when $h=0$ in position space.
 
Because $\tilde{G}^{\rm power}$ is $C^\infty$ away from $\kappa=0$, forming the ratio $\tilde{G}^{\rm sparse}_\kappa/\tilde{G}^{\rm power}_\kappa$ doesn't affect the local smoothness properties of $\tilde{G}^{\rm power}_\kappa$.  However, this ratio does cancel out part of the overall trend whereby $\tilde{G}^{\rm sparse}_\kappa$ gets bigger near $\kappa=0$ and $\kappa=L$.  As a result, studying $\tilde{G}^{\rm sparse}_\kappa/\tilde{G}^{\rm power}_\kappa$ rather than $\tilde{G}^{\rm sparse}_\kappa$ by itself makes it easier to accurately pick out the local smoothness properties from a finite sampling of points.  As on the $2$-adic side, the numerical data approximately follow exponential trajectories:
 \eqn{archDeltaLaw}{
  \tilde{A}^{\rm power}(N,s) \approx s(N-1) + \log_2 K^{\rm power}(s) \,,
 }
where $K(s)$ is $N$-independent.  These trajectories persist even at positive $s$, after Archimedean continuity is lost.  So we can usefully define
 \eqn{alphatArch}{
  \tilde\alpha^{\rm power}(N,s) \equiv -\tilde{A}^{\rm power}(N,s) + 
    \tilde{A}^{\rm power}(N-1,s) \,,
 }
and then $\tilde\alpha^{\rm power}(N,s) \approx -s$ for large $N$ is our numerical estimate of the best (i.e.~most positive) exponent appearing in a local Archimedean H\"older condition for $\tilde{G}^{\rm sparse}_\kappa$.

\subsection{Local smoothness in position space}
\label{SmoothnessPosition}

Position space smoothness can be studied using quantities analogous to the ones used in section~\ref{SmoothnessMomentum} for momentum space.  Specifically, we define
 \eqn{AdefX}{
  A^{2-\rm adic}(N,s) &\equiv \log_2 \max_{h\ \rm odd} 
     \left| {G^{\rm sparse}_h \over G^{2-\rm adic}_h} - 
       {G^{\rm sparse}_{h+L/2} \over G^{2-\rm adic}_{h+L/2}} 
     \right|_\infty  \cr
  \alpha^{2-\rm adic}(N,s) &\equiv -A^{2-\rm adic}(N,s) + A^{2-\rm adic}(N-1,s) \,,
 }
and then, assuming $\alpha^{2-\rm adic}(N,s)$ is nearly constant for large $N$, its large $N$ limit is our numerical estimate of the best possible local H\"older exponent for $G^{\rm sparse}_h$ in a $2$-adic setting.  Likewise, we define
 \eqn{AdefXArch}{
  A^{\rm power}(N,s) &\equiv \log_2 \max_{{L \over 4} \leq h < {3L \over 4}}
    \left| {G^{\rm sparse}_h \over G^{\rm power}_h} - 
       {G^{\rm sparse}_{h+1} \over G^{\rm power}_{h+1}} 
     \right|_\infty  \cr
  \alpha^{\rm power}(N,s) &\equiv -A^{\rm power}(N,s) + A^{\rm power}(N-1,s) \,.
 }
The large $N$ limit of $\alpha^{\rm power}(N,s)$ (assuming it exists) is our numerical estimate of the best possible local H\"older exponent for $G^{\rm sparse}_h$ in an Archimedean setting.
 \CaptionedFigure[6.5in]{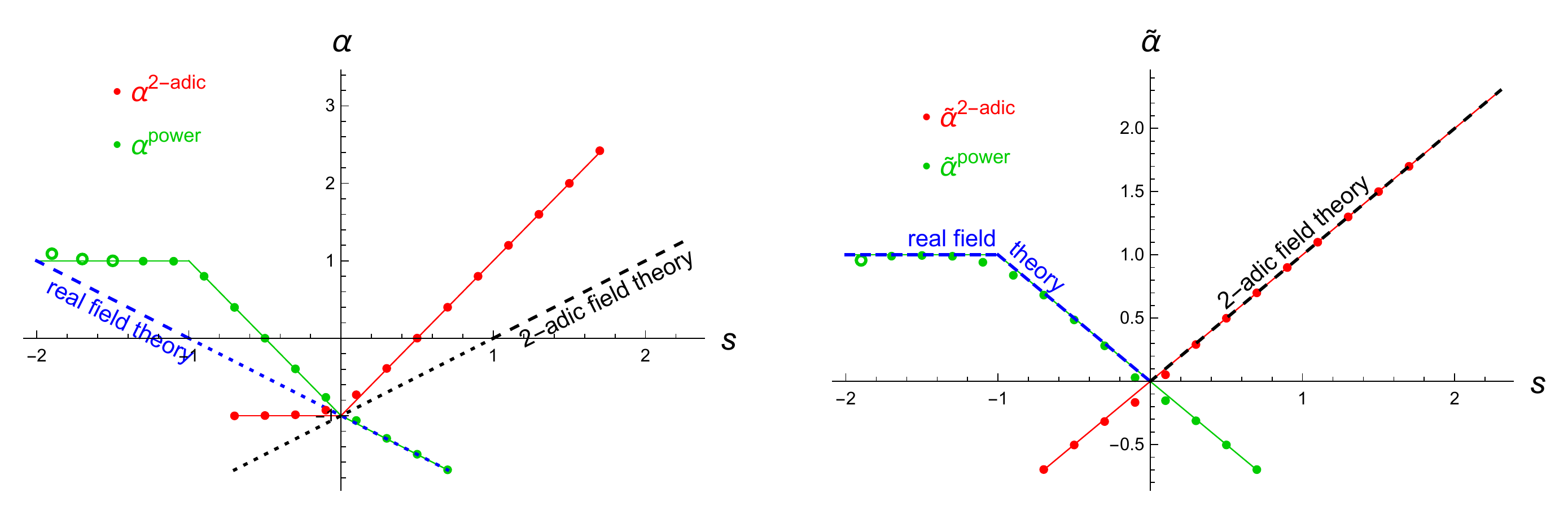}{
  $\alpha$ versus $s$ and $\tilde\alpha$ versus $s$ in the $2$-adic and Archimedean settings.  Field theory bounds derived in section~\ref{CONTINUITY} are shown in dashed black and dashed blue.  Dotted black and dotted blue show the naive extrapolations of these bounds to negative $\alpha$ and $\tilde\alpha$.  Red and green dots are numerical evaluations of $\alpha$ and $\tilde\alpha$ as defined in sections~\ref{SmoothnessPosition} and~\ref{SmoothnessMomentum}, respectively, with $N=20$.  Solid red and green lines show the obvious piecewise linear trends which approximately match the numerical evaluations.  Open circles denote evaluations in which we restricted ${7L \over 16} \leq h < {9L \over 16}$; otherwise we use half the available points as explained in the main text.  For $s\leq -2$, convergence of the sparse model to the nearest neighbor model implies that $\alpha=\tilde\alpha=1$, but our numerical scheme for picking out $\alpha$ and $\tilde\alpha$ becomes less reliable in this region due to difficulty normalizing $G^{\rm sparse}$ and $G^{\rm power}$ in a mutually consistent way.}

We find good evidence that $\alpha^{2-\rm adic}(N,s)$ and $\alpha^{\rm power}(N,s)$ have finite large $N$ limits.  Our numerical results are well described by piecewise linear dependence of $\alpha$ on $s$, and in particular by
 \eqn{alphaDesc}{\seqalign{\span\TL & \span\TR &\qquad \span\TT\quad & \span\TL & \span\TR}{
  \alpha^{\rm power} &= -2(s+1/2) & for & -1 &< s<0  \cr
  \alpha^{2-\rm adic} &= 2(s-1/2) & for & 0 &< s<1 \,.
 }}
See figure~\ref{AlphaVsS}.  Two caveats on our numerical results can be summarized as follows:
 \begin{itemize}
  \item When $|s| > 1$, it becomes harder to get good numerical results, particularly on the Archimedean side, because the functions under consideration are quite smooth, and we have to compute very small differences accurately.  Even apart from issues of numerical accuracy, it becomes challenging on the Archimedean side to distinguish between rapid but smooth variation and the slightly non-smooth behavior that determines the H\"older exponent.
  \item Numerical evaluations of H\"older exponents diverge a bit from expectations at $s=0$, and also at $s=-1$.  This is not too surprising, given that our estimates of the prefactors $K$ in the H\"older inequalities show divergences at these values of $s$: See for example \eno{Kvalue}, \eno{psiDiffEst}, and \eno{GxEstReal}.  Possibly at these special values we need logarithmic corrections to the relevant H\"older condition.  It is also possible that simple piecewise linear functions only approximately fit the dependence of $\alpha$ on $s$.  More extensive and accurate numerical investigations are needed in order to establish fully reliable results.
 \end{itemize}

\subsection{Transition between two types of smoothness}
\label{Transition}

The most interesting regime in position space is $-1<s<1$, where we are losing Archimedean continuity and gaining $2$-adic continuity.  We focus in this section entirely on this regime, and we present the simplest account of the transition from Archimedean to ultrametric continuity which is consistent with our numerics.  Due to finite numerical resolution, we cannot rigorously determine the measure-theoretic behavior of the position space Green's functions in regions where the Green's functions are very ragged.  We attempt to qualify our claims below with the appropriate level of confidence.

In momentum space, our numerics are consistent with there being a single exponent on the $2$-adic side, $\tilde\alpha^{2-\rm adic} = s$, which describes both the global H\"older continuity condition over all $k$ and the local continuity at each possible value of $k$.  In other words, as far as we can tell, the function $\tilde{G}(k)$ is equally ragged everywhere.  A similarly uniform story applies on the Archimedean side, with $\tilde\alpha^{\rm power} = -s$.  Numerical results are fully in accord with expectations from field theory, where we were able to compute $\tilde\alpha^{2-\rm adic}$ and $\tilde\alpha^{\rm power}$ analytically.  The upshot is that the transition from Archimedean to ultrametric continuity happens rather simply, with ordinary continuity failing just as $2$-adic continuity emerges: i.e.~$\tilde\alpha^{\rm power}$ becomes negative just as $\tilde\alpha^{2-\rm adic}$ becomes positive, at $s=0$.

The field theory estimates of the H\"older exponents for the position-space Green's function were $s-1$ on the $2$-adic side and $-s-1$ on the Archimedean side.  These exponents (uniformly negative in the $-1<s<1$) were based entirely on the average scaling of $\tilde{G}(k)$ as a power of $|k|$ far from $k=0$.  As such, they tell us the global H\"older exponent, which we believe characterizes the behavior of $G(x)$ close to $x=0$: That is, $G(x) \approx |x|_2^{s-1}$ on the $2$-adic side, while $G(x) \approx |x|_\infty^{-s-1}$ on the Archimedean side.  The surprise we get from numerics is that away from $x=0$, a more complicated dependence of H\"older smoothness on $s$ emerges, with local H\"older exponents $\alpha$ somewhat more positive than the field theory bounds: That is, $G(x)$ seems to be somewhat smoother away from the origin than its behavior right near $x=0$.  Our numerical results are consistent with there being a piecewise linear dependence of $\alpha$ on $s$, as summarized in particular by \eno{alphaDesc}.  These results \eno{alphaDesc} indicate that Archimedean continuity of $G^{\rm sparse}_h$ is lost at $s=-1/2$, but $2$-adic continuity doesn't emerge until $s=1/2$.  We may well ask, what happens for $-1/2<s<1/2$, when both $\alpha^{\rm power}$ and $\alpha^{2-\rm adic}$ are negative?

To better understand the region of transition between the Archimedean and 2-adic smoothness,  it is instructive to inspect overlaid plots of the Green's function for different system sizes, see figures \ref{padicSpikes2} and \ref{ArchimedeanSpikes}.
 \CaptionedFigure[6.5in]{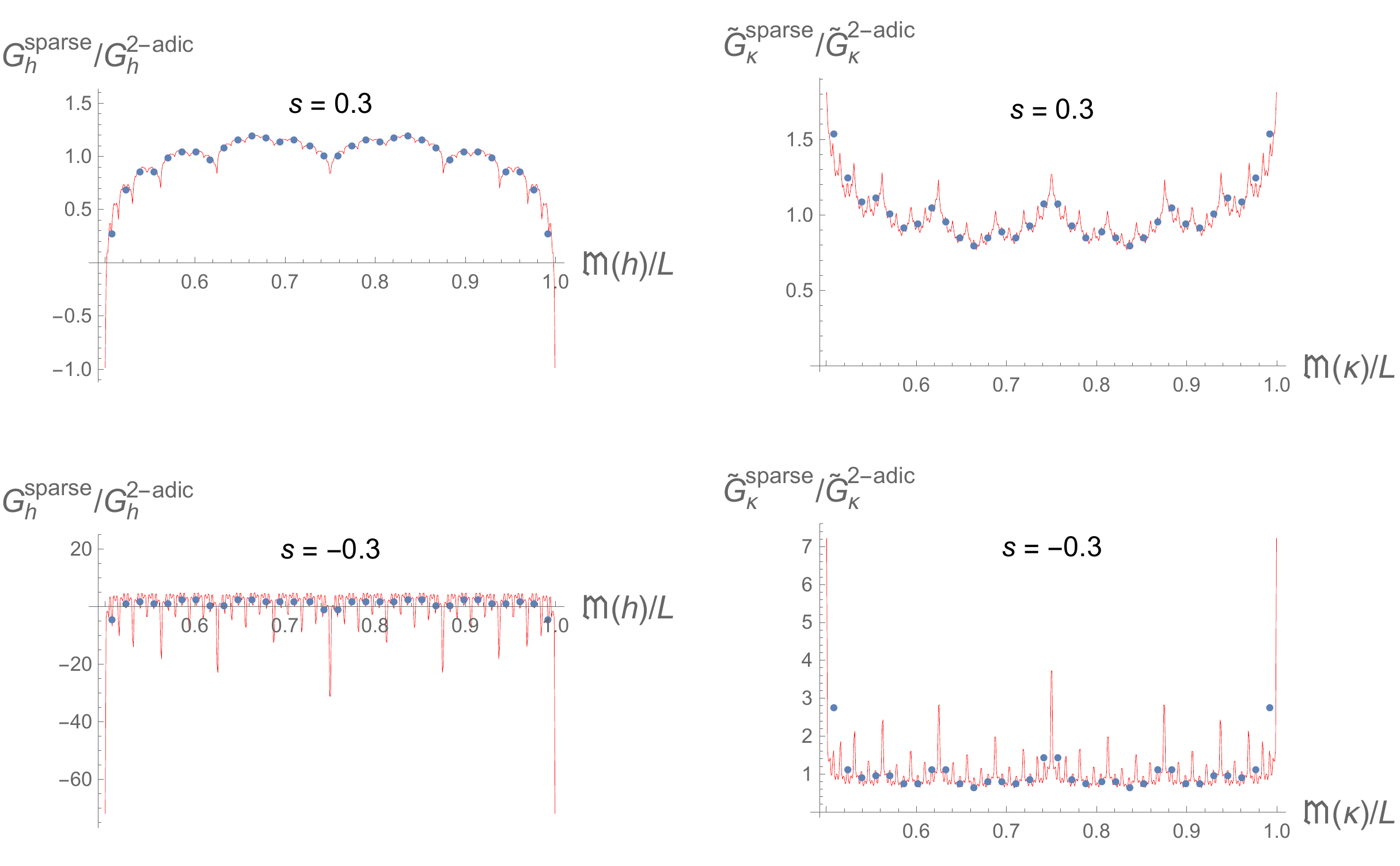}{Plots of $G^{\rm sparse}_h / G^{2-\rm adic}_h$ and $\tilde{G}^{\rm sparse}_\kappa / \tilde{G}^{\rm 2-adic}_\kappa$ over the Monna map of the odd integers. As $s$ becomes more positive, the numerical data is closer to a $2$-adically continuous curve when $N$ is large.  Blue points are for $N=6$, while the red curves are for $N=10$.}

 \CaptionedFigure[6.5in]{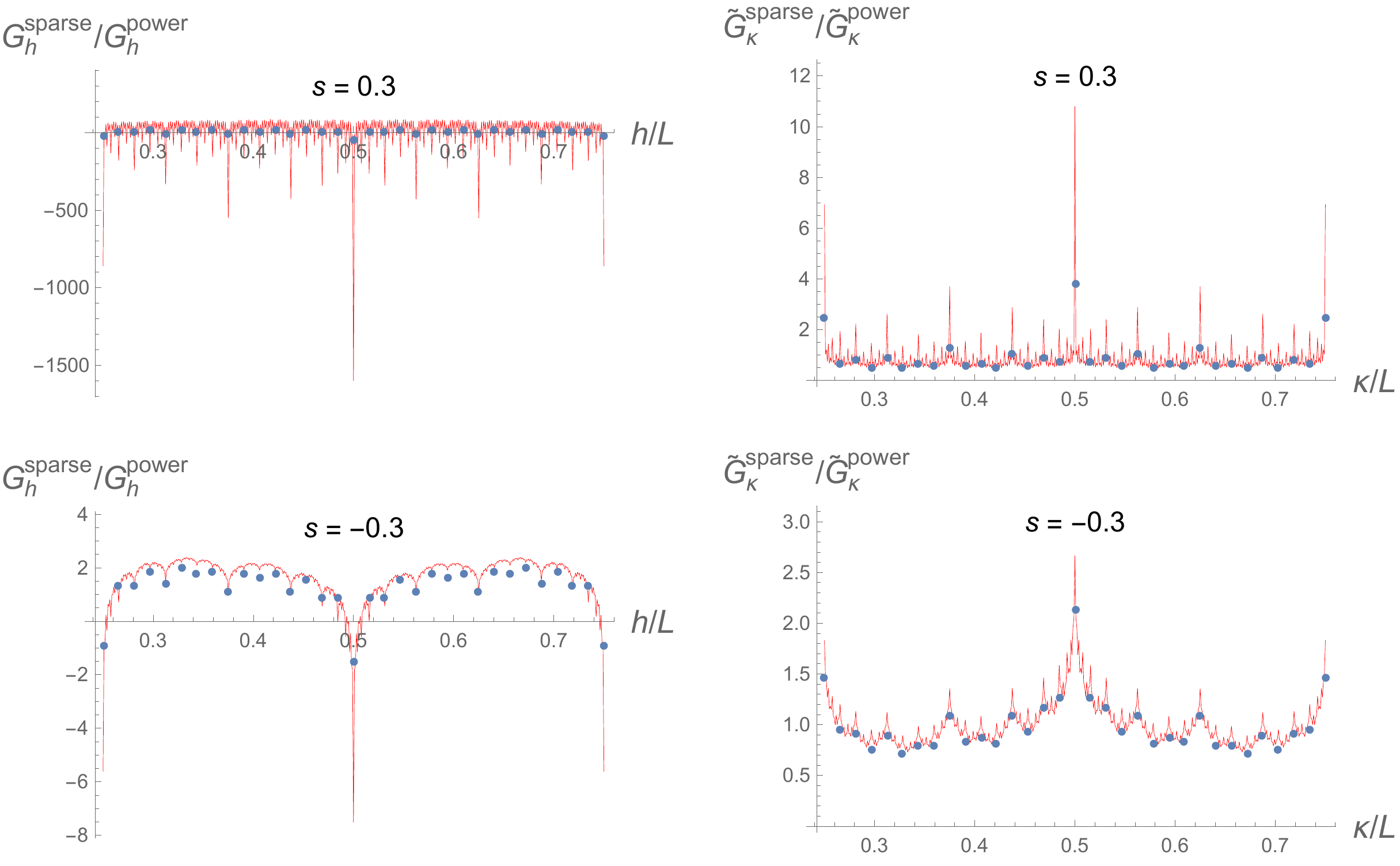}{Plots of $G^{\rm sparse}_h / G^{\rm power}_h$ and $\tilde{G}^{\rm sparse}_\kappa / \tilde{G}^{\rm power}_\kappa$ over the middle half of points. As $s$ becomes more negative, the numerical data is closer to a continuous curve when $N$ is large.  Blue points are for $N=6$, while the red curves are for $N=10$.}

Based on these figures and related studies, the scenario we regard as most likely is that for $-1/2<s<0$, the continuum limit of $G^{\rm power}_h$ defines an absolutely continuous measure, $G(x) dx$, with respect to ordinary Lebesgue measure $dx$, but for $s>0$ any such continuum limit would necessarily have a singular term in its Radon-Nikodym decomposition.  Similarly, we suggest that for $0<s<1/2$, the continuum limit of $G^{2-\rm adic}_h$ defines an absolutely continuous measure with respect to the standard Haar measure on $\mathbb{Q}_2$ while for $s<0$ any such continuum limit would have a singular term (with respect to the Haar measure on $\mathbb{Q}_2$) in its Radon-Nikodym decomposition. We find support for the claim of absolutely continuous measures in the above-mentioned regimes when we study the scaling of the height of the spikes in figures ~\ref{padicSpikes2} and~\ref{ArchimedeanSpikes} as a function of $N$: the weight of each spike (meaning the integral over a small region including the spike) distinctly appears to tend to zero with increasing $N$.
When singular terms in Radon-Nikodym decompositions do exist, we conjecture that they have as their support sets which are dense in position space.

One way in which singular terms in Radon-Nikodym decompositions could arise is for the continuum limit $G(x)$ to include delta functions.  Inspection of figure~\ref{padicSpikes2} is consistent with there being a dense set of delta function spikes in $G(x)$ as a function of $2$-adic $x$ when $s=-0.3$, but none when $s=0.3$.  Similarly, figure~\ref{ArchimedeanSpikes} is consistent with there being a dense set of delta function spikes in $G(x)$ as a function of real $x$ when $s=0.3$, but none with $s=-0.3$.  The discerning reader may note, however, that the spikes on the Archimedean side are stronger at $s=0.3$ than the ones on the $2$-adic side at $s=-0.3$.  This asymmetry manifests itself in the scaling of the height of these spikes with $N$, for the weight of each spike grows with $N$ on the Archimedean side for $s=0.3$, but may be trending very slowly toward $0$ on the $2$-adic side at $s=-0.3$.  A related effect appears in figure~\ref{AlphaVsS}: $\alpha^{2-\rm adic} \approx -1$ for $s<0$, while $\alpha^{\rm power} \approx -1-s$ for $s>0$. 

Inspection of figures~\ref{padicSpikes2} and~\ref{ArchimedeanSpikes} reveals some self-similarity in the Green's functions both before and after the Monna map is applied.  We have not investigated this fractal behavior in detail; however, we note that similar behavior has been found independently in band structure calculations in connection with proposed cold atom experiments \cite{SchleierSmithDiscussions}.

\section{Discussion}

For decades, $p$-adic numbers have been considered as an alternative to real numbers as a notion of continuum which could underlie fundamental physics at a microscopic scale; see for example \cite{Vladimirov:1994zz}.  The current study shows how the large system size limit of an underlying discrete system naturally interpolates between a one-dimensional Archimedean continuum and a $2$-adic continuum as we vary a spectral exponent.  By focusing a free field example, we are able to solve the model through essentially trivial Fourier space manipulations.  The correlators of the theories we study are all determined in terms of the two-point function through application of Wick's theorem.  The two-point function is smooth in an Archimedean sense when $s$ is sufficiently negative, and in a $2$-adic sense when $s$ is sufficiently positive.  The transition from these two incompatible notions of continuity can be precisely characterized in terms of H\"older exponents characterizing the smoothness of the two-point function.  We have found the dependence of these exponents on $s$ through a combination of analytical field theory arguments and numerics on finite but large systems.

\begin{figure}
  \centerline{\begin{tikzpicture}
   %Colors
   \colorlet{darkgreen}{green!80!black}
   %Coordinates
   \tikzmath{
     \hscale = 5;
     \pointsize = 0.1;
     \p1 = \hscale;
     \p4 = 0.5 * \hscale;
     \p6 = 0.67 * \hscale;
     \a1 = -\hscale;
     \a4 = -0.5 * \hscale;
     \a6 = -0.67 * \hscale;
    }
   \coordinate (C) at (0,0);
   \coordinate (P1) at (\p1,0);
   \coordinate (P4) at (\p4,0);
   \coordinate (P6) at (\p6,0);
   \coordinate (RP4) at (\p1,\p4);
   \coordinate (N4) at (1.1*\p4,0.7*\p4);
   \coordinate (RP6) at (\p1,0.33*\hscale);
   \coordinate (N6) at (\p6,0.3*\p6);
   \coordinate (A1) at (\a1,0);
   \coordinate (A4) at (\a4,0);
   \coordinate (A6) at (\a6,0);
   \coordinate (RA4) at (\a1,\p4);
   \coordinate (M4) at (1.1*\a4,0.7*\p4);
   \coordinate (RA6) at (\a1,0.33*\hscale);
   \coordinate (M6) at (\a6,0.3*\p6);
   \coordinate (V1) at (0,0.6*\hscale);
   \coordinate (V2) at (0,-0.35*\hscale);
   \coordinate (B1) at (1.15*\p1,0);
   %Draw the main line and the points
   \draw [dashed] (V1) -- (V2);
   \draw [line width=4pt,darkgreen] (A4) -- (P4);
   \draw [->, line width=2pt] (A1) -- (B1);
   \draw [fill] (C) circle [radius=\pointsize];
   \draw [fill] (P1) circle [radius=\pointsize];
   \draw [fill] (A1) circle [radius=\pointsize];
   %Label the points.  right=1pt means to the right by the standard length plus 1pt
   \node [below=5pt] at (C) {$s=0$};
   \node [below=5pt] at (P1) {$1$};
   \node [below=5pt] at (P4) {$1 \over 2$};
   \node [below=5pt] at (P6) {$2 \over 3$};
   \node [below=5pt] at (A1) {$-1$};
   \node [below=5pt] at (A4) {$-{1 \over 2}$};
   \node [below=5pt] at (A6) {$-{2 \over 3}$};
   %Draw conjectured RG fixed points on ultrametric side
   \draw [fill,red] (P4) circle [radius=\pointsize];
   \draw [line width=2pt,red] (P4) arc (180:90:0.5*\hscale);
   \node [rotate=60,red] at (N4) {$\phi^4$ relevant};
   \draw [fill,red] (RP4) circle [radius=\pointsize];
   \draw [fill,blue] (P6) circle [radius=\pointsize];
   \draw [line width=2pt,blue] (P6) arc (180:90:0.33*\hscale);
   \node [rotate=60,blue] at (N6) {$\phi^6$ relevant};
   \draw [fill,blue] (RP6) circle [radius=\pointsize];
   %Draw conjectured RG fixed points on Archimedean side
   \draw [fill,red] (A4) circle [radius=\pointsize];
   \draw [line width=2pt,red] (A4) arc (0:90:0.5*\hscale);
   \node [rotate=-60,red] at (M4) {$\phi^4$ relevant};
   \draw [fill,red] (RA4) circle [radius=\pointsize];
   \draw [fill,blue] (A6) circle [radius=\pointsize];
   \draw [line width=2pt,blue] (A6) arc (0:90:0.33*\hscale);
   \node [rotate=-60,blue] at (M6) {$\phi^6$ relevant};
   \draw [fill,blue] (RA6) circle [radius=\pointsize];
   %Hint at further branches of RG fixed points
   \draw [fill] (\p6+0.2*\hscale,0.23*\hscale) circle [radius=0.5*\pointsize];
   \draw [fill] (\p6+0.25*\hscale,0.205*\hscale) circle [radius=0.5*\pointsize];
   \draw [fill] (\p6+0.3*\hscale,0.18*\hscale) circle [radius=0.5*\pointsize];
   \draw [fill] (\a6-0.2*\hscale,0.23*\hscale) circle [radius=0.5*\pointsize];
   \draw [fill] (\a6-0.25*\hscale,0.205*\hscale) circle [radius=0.5*\pointsize];
   \draw [fill] (\a6-0.3*\hscale,0.18*\hscale) circle [radius=0.5*\pointsize];
   %Further labeling
   \node [above=5pt,darkgreen] at (C) {\small No relevant deformations};
   \node [below=30pt,black] at (P6) {ULTRAMETRIC};
   \node [below=30pt,black] at (A6) {ARCHIMEDEAN};
   \node [above=10pt,right=-10pt] at (A1) {Gaussian};
   \node [above=10pt,left=-10pt] at (P1) {Gaussian};
   \node [right] at (B1) {$s$};
  \end{tikzpicture}}
  \caption{Conjectured pattern of fixed points of the renormalization group for interacting field theories of a single bosonic scalar field with $\phi \to -\phi$ symmetry.}\label{Deformations}
 \end{figure}
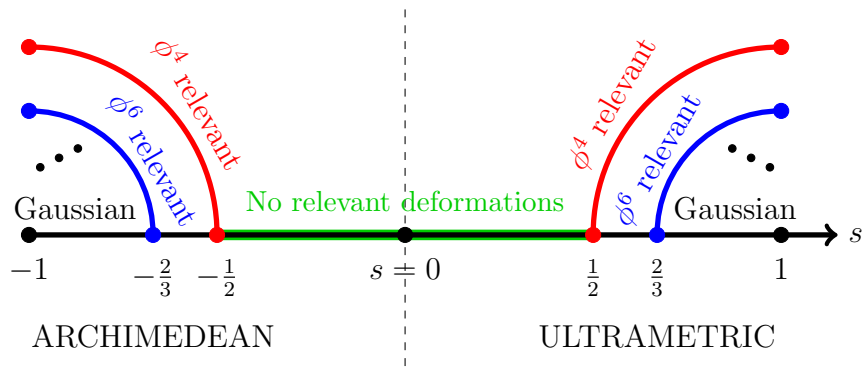

Quite a wide range of generalizations of our basic construction can be contemplated:
 \begin{enumerate}
  \item We can generalize to primes $p>2$.  One significant subtlety arises when doing so, namely the structure within $\mathbb{Z}/p\mathbb{Z}$ of sparse couplings.  The simplest alternative is for spin $0$ to couple to spins $\pm\theta p^n$ with a strength $p^{ns}$, where $\theta$ runs over all elements of $\{1,2,3,\dots,p-1\}$.  This coupling pattern is featureless within $\mathbb{Z}/p\mathbb{Z}$ because it treats all values of $\theta$ the same.  We could however contemplate other possibilities.  For example, if $p=5$, an interesting alternative is to introduce couplings only for $\theta=1$ and $\theta=4$ (the quadratic residues).  More generally, one could expand the dependence of couplings on $\theta$ in a sum of multiplicative characters over $\mathbb{Z}/p\mathbb{Z}$.
  \item We focused entirely on bosonic spins $\phi_i$, but there is no reason not to consider fermions $c_i$ instead.  Then the coupling matrix $J_{ij}$ would have to be anti-symmetric, and likewise the two-point Green's function would be odd.  Within this framework one could consider a variety of sparse coupling patterns.
  \item Higher-dimensional examples are not hard to come by.  Consider real bosonic spins $\phi_{\vec\imath}$ labeled by a two-dimensional vector $\vec\imath = i_1 \hat{1} + i_2 \hat{2}$, where $i_1$ and $i_2$ are in $\mathbb{Z}/3^N\mathbb{Z}$.  Suppose we establish a coupling matrix $J_{\vec\imath\,\vec\jmath} = J_{\vec\imath-\vec\jmath}$ where
 \eqn{Jthree}{
  J_{\vec{h}} = \left\{ 
   \seqalign{\span\TR &\qquad\span\TT\qquad & \span\TR &\qquad\span\TT\qquad & \span\TR}{
   3^{\min\{n_1,n_2\} s} & if & h_1 = \pm 3^{n_1} & and & h_2 = \pm 3^{n_2}  \cr
   3^{n_1 s} & if & h_1=0 & and & h_2 = \pm 3^{n_2}  \cr
   3^{n_2 s} & if & h_2=0 & and & h_1 = \pm 3^{n_1} \,,
    } \right.
 }
with all other entries vanishing except $J_0$, whose value we choose in order to have the Fourier coefficient $\tilde{J}_{\vec\kappa}$ vanish when $\vec\kappa = 0$.  Then for sufficiently negative $s$ we have effectively a nearest neighbor model which approximates the massless field theory $S = \int d^2 x \, {1 \over 2} (\nabla\phi)^2$.  For $s$ sufficiently positive, one obtains instead a continuum theory over $\mathbb{Z}_3 \times \mathbb{Z}_3$, which can be understood as the ring of integers in the unramified quadratic extension of $\mathbb{Q}_3$.
 \end{enumerate}
All the examples above remain within the paradigm of free field theory.  Still easy to formulate, but obviously much harder to solve, are interacting theories with sparse couplings.  For example, we could start with any of the models introduced in section~\ref{SETUP} and add a term $\sum_i V(\phi_i)$ to the Hamiltonian describing arbitrary on-site interactions.  To get some first hints of what to expect these interactions to do, recall in $2$-adic field theory that $G(x) \approx |x|_2^{s-1}$ at small $x$.  Comparing this to the standard expectation $G(x) \approx |x|_2^{2\Delta_\phi}$, we arrive at $\Delta_\phi = (1-s)/2$ as the ultraviolet dimension of $\phi$.  When describing perturbations of the Gaussian theory, we can use normal UV power counting: $[\phi^n] = n\Delta_\phi$.  Thus $\phi^n$ is relevant when $s > 1-2/n$.  If we impose $\mathbb{Z}_2$ symmetry, $\phi \to -\phi$, then in the region $s<1/2$, the Gaussian theory has no relevant local perturbations, but as $s$ increases from $1/2$ to $1$, first $\phi^4$ and then higher powers of $\phi^2$ become relevant.  It is reasonable to expect some analog of Wilson-Fisher fixed points to appear.  Possibly as $s \to 1$ these fixed points extrapolate to analogs of minimal models.  An analogous story presumably applies on the Archimedean side to power-law field theories controlled by $s$ in the range $(-1,0)$, with $G(x) \approx |x|_\infty^{-s-1}$ and therefore $\Delta\phi = (1+s)/2$.  See figure~\ref{Deformations}.

The sparse coupling theories are sufficiently similar to $2$-adic field theories for $s>0$ and to power-law field theories for $s<0$ that it is reasonable to conjecture that the same pattern of renormalization group fixed points arises.  This line of reasoning leaves out a lot, though: In particular, we have no deep understanding of how the improved local H\"older smoothness arises, nor how it might affect renormalization group flows.  A Monte Carlo study of the phases of the sparsely coupling Ising model might help refine our understanding of the renormalization group flows available to interacting models, particularly in the range $-2/3 < s < 2/3$ where no powers of $\phi$ higher than $\phi^4$ are relevant---according at least to naive power counting as presented here.

\subsection*{Acknowledgments}

We thank S.~Hartnoll for getting us started on this project by putting us in touch with M.~Schleier-Smith's group, and we particularly thank G.~Bentsen and M.~Schleier-Smith for extensive discussions.  This work was supported in part by the Department of Energy under Grant No.~DE-FG02-91ER40671, and by the Simons Foundation, Grant 511167 (SSG).  The work of C.~Jepsen was supported in part by the National Science Foundation under Grant No.~PHY-1620059.

\clearpage

\bibliographystyle{ssg}
\bibliography{pattern}

\begingroup\raggedright\begin{thebibliography}{1}

\bibitem{Gouvea:1997zz}
F.~Q. Gouv{\^e}a, {\em p-adic Numbers}.
\newblock Springer, 1997.

\bibitem{Dyson:1968up}
F.~J. Dyson, ``{Existence of a phase transition in a one-dimensional Ising
  ferromagnet},'' {\em Commun. Math. Phys.} {\bf 12} (1969) 91--107.

\bibitem{Bleher:1973zz}
P.~M. Bleher and J.~G. Sinai, ``Investigation of the critical point in models
  of the type of Dyson's hierarchical models,'' {\em Comm. Math. Phys.} {\bf
  33} (1973), no.~1 23--42.

\bibitem{Lerner:1989ty}
E.~{\relax Yu}. Lerner and M.~D. Missarov, ``{Scalar Models of $p$-adic Quantum
  Field Theory and Hierarchical Models},'' {\em Theor. Math. Phys.} {\bf 78}
  (1989) 177--184.

\bibitem{Missarov:2012zz}
M.~Missarov, ``$p$-Adic Renormalization Group Solutions and the Euclidean
  Renormalization Group Conjectures,'' {\em $P$-Adic Numbers, Ultrametric
  Analysis, and Applications} {\bf 4} (2012), no.~2 109--114.

\bibitem{SchleierSmithDiscussions}
{G.~Bentsen, E.~Davis, and M.~Schleier-Smith, private discussions}.

\bibitem{Sally:1998zz}
P.~J. Sally, ``An Introduction to $p$-adic Fields, Harmonic Analysis and the
  Representation Theory of $SL(2)$,'' {\em Letters in Mathematical Physics}
  {\bf 46} (Oct, 1998) 1--47.

\bibitem{Hung:2016zz}
C.-L. {Hung}, A.~{Gonz{\'a}lez-Tudela}, J.~I. {Cirac}, and H.~J. {Kimble},
  ``{Quantum spin dynamics with pairwise-tunable, long-range interactions},''
  {\em Proceedings of the National Academy of Science} {\bf 113} (Aug., 2016)
  E4946--E4955, \href{http://arxiv.org/abs/1603.05860}{{\tt 1603.05860}}.

\bibitem{Vladimirov:1994zz}
V.~Vladimirov, I.~Volovich, and E.~Zelenov, {\em P-adic Analysis and
  Mathematical Physics}.
\newblock Series on Soviet and East European mathematics. World Scientific,
  1994.

\end{thebibliography}\endgroup

\end{document}